\documentclass[letterpaper,english,pra, superscriptaddress, twocolumn, nofootinbib]{revtex4-1}

\usepackage[T1]{fontenc}
\usepackage{color}
\usepackage{babel}
\usepackage{pifont}
\usepackage{amsmath}
\usepackage{amssymb}
\usepackage{graphicx}
\usepackage[unicode=true,pdfusetitle,
 bookmarks=true,bookmarksnumbered=false,bookmarksopen=false,
 breaklinks=false,pdfborder={0 0 1},backref=false,colorlinks=true]
 {hyperref}
\hypersetup{
 allcolors=blue}

\makeatletter

\pdfpageheight\paperheight
\pdfpagewidth\paperwidth

\AtBeginDocument{
  
}

\makeatother

\begin{document}

\title{The magnetic dipole-dipole interaction induced by electromagnetic
field}

\author{Jiaxuan Wang}

\affiliation{Texas A\&M University, College Station, TX 77843, USA}

\affiliation{University of Science and Technology, Hefei 230026, China}

\author{Hui Dong}

\affiliation{Graduate School of Chinese Academy of Engineering Physics, Beijing,
China}

\author{Sheng-Wen Li}
\email{lishengwen@tamu.edu}

\affiliation{Texas A\&M University, College Station, TX 77843, USA}

\affiliation{Baylor University, Waco, TX 76798}

\date{\today}
\begin{abstract}
We give a derivation for the indirect interaction between two magnetic
dipoles induced by the quantized electromagnetic field. It turns out
that the interaction between permanent dipoles directly returns to
the classical form; the interaction between transition dipoles does
not directly return to the classical result, yet returns in the short-distance
limit. In a finite volume, the field modes are highly discrete, and
both the permanent and transition dipole-dipole interactions are changed.
For transition dipoles, the changing mechanism is similar with the
Purcell effect, since only a few number of nearly resonant modes take
effect in the interaction mediation; for permanent dipoles, the correction
comes from the boundary effect: if the dipoles are placed close to
the boundary, the influence is strong, otherwise, their interaction
does not change too much from the free space case.
\end{abstract}
\maketitle

\section{Introduction}

The interaction between particles is induced by their local interaction
with the field. This is a basic understanding in modern physics, and
should also applies for the interaction between two electric/magnetic
dipoles. Thus, by controlling the property of the electromagnetic
(EM) field, one can artificially engineer the dipole-dipole interaction
\cite{liao_single-photon_2014,liao_single-photon_2015,lambert_cavity-mediated_2016,baranov_magnetic_2016,liao_dynamical_2016,shahmoon_nonradiative_2013,shahmoon_dispersion_2013,cai_symmetry-protected_2016,shahmoon_casimir_2017,donaire_dipole-dipole_2017,cortes_super-coulombic_2017},
which widely appears in many different microscopic systems, such as
the interaction between the Josephson qubit and the dielectric defects
\cite{martinis_decoherence_2005,paik_observation_2011,rigetti_superconducting_2012,lisenfeld_decoherence_2016},
the interaction between the nitrogen-vacancy and the nuclear spins
around \cite{doherty_nitrogen-vacancy_2013,zhao_decoherence_2012},
as well as the dipoles in chemical and biology molecular \cite{yang_dimerization-assisted_2010,el-ganainy_resonant_2013,dong_photon-blockade_2016}. 

In classical electrodynamics, the interactions between two electric/magnetic
dipoles are given by \cite{jackson_classical_1998}
\begin{align}
V_{\mathrm{e}} & =\frac{1}{4\pi\epsilon_{0}}\frac{\vec{p}_{1}\cdot\vec{p}_{2}-3(\vec{p}_{1}\cdot\hat{\mathrm{e}}_{\mathbf{r}})(\vec{p}_{2}\cdot\hat{\mathrm{e}}_{\mathbf{r}})}{r^{3}},\nonumber \\
V_{\mathrm{m}} & =\frac{\mu_{0}}{4\pi}\frac{\vec{m}_{1}\cdot\vec{m}_{2}-3(\vec{m}_{1}\cdot\hat{\mathrm{e}}_{\mathbf{r}})(\vec{m}_{2}\cdot\hat{\mathrm{e}}_{\mathbf{r}})}{r^{3}}.\label{eq:classical}
\end{align}
 Thus it is natural to expect such interaction can be derived using
quantum mechanics, based on the idea of the mediation of the quantized
EM field. 

The field induced interaction between two electric dipoles has been
studied based on both the Heisenberg equation \cite{lyuboshitz_resonance_1968,lehmberg_radiation_1970,ficek_quantum_1987}
and the master equation \cite{agarwal_quantum_1974,ficek_quantum_2005}.
In these studies, two resonant electric dipoles with the same transition
frequency are concerned, and an interaction Hamiltonian $\hat{H}_{\mathrm{e}}=\xi(\hat{\sigma}_{1}^{+}\hat{\sigma}_{2}^{-}+\hat{\sigma}_{1}^{-}\hat{\sigma}_{2}^{+})$
is derived from the mediation of the field. The interaction strength
$\xi$ does not directly return to the above classical result, but
returns in the short-distance limit $r/\lambda\ll1$ \cite{dicke_coherence_1954},
where $r$ is the distance between the two dipoles, and $\lambda$
is the wavelength of the transition frequency.

Notice that there is also some conceptual difficulty when studying
the electric dipole interaction induced by the EM field, i.e., the
static electric interaction is induced by the longitudinal modes of
the EM field, which are not quantized in the Coulomb gauge \cite{agarwal_quantum_1974,cohen-tannoudji_photons_1989}.
If the Lorenz gauge is adopted, some other conceptual difficulties,
e.g., the negative probability problem, also arise \cite{ryder_quantum_1996},
which makes it uneasy to get a clear picture on this problem.

On the contrast, the magnetic interaction only involves the transverse
modes of the EM field, which can be well quantized under the Coulomb
gauge, thus it could be clear to study the magnetic dipole-dipole
interaction \cite{baranov_magnetic_2016,lambert_cavity-mediated_2016}.
In this paper, we give a simple derivation for this indirect interaction
between two magnetic dipoles induced by the EM field. Our derivation
goes through the following procedure:

1) First, only dipole-1 is put in the EM field, and that generates
a dipole field.

2) The magnetic field contains both the vacuum field and the dipole
field, and the interaction between dipole-2 and the dipole field leads
to the dipole-dipole interaction.

Based on this idea, we obtain an interaction Hamiltonian for the two
magnetic dipoles, which is formally exact and naturally has a retarded
structure. After proper Markovian approximation and rotating-wave
approximation (RWA), the interaction reduces to a time-local one.

Here we concern both the \emph{permanent dipole }and \emph{transition
dipole}, which correspond to the diagonal and off-diagonal elements
of the dipole operator respectively. Our result shows that, in free
space, the interaction between the permanent dipoles directly returns
to the classical interaction; the interaction between transition dipoles
has the same form with the previous studies on electric dipole interaction
\cite{agarwal_quantum_1974,lehmberg_radiation_1970,ficek_quantum_1987},
and it does not directly return to the classical result, but returns
in the limit $r/\lambda\ll1$.

We also study the dipole-dipole interaction in a finite volume, where
the field modes are highly discrete. Both the permanent and transition
dipole-dipole interactions are changed from the free space case, but
by different mechanisms. For transition dipoles, this changing mechanism
is similar with the Purcell effect \cite{purcell_resonance_1946,scully_quantum_1997},
since only a few number of nearly resonant modes take effect in the
mediation of the interaction; for permanent dipoles, still all the
field modes take effect for the interaction mediation, and the correction
comes from the boundary effect: if the dipoles are placed close to
the boundary, the influence is strong, if they are both placed far
away from the boundary, their interaction does not change too much
from the free space case, and this is also similar with the situation
in classical electrodynamics.

The paper is arranged as follows: in Sec.\,II, we derive the retarded
dipole-dipole interaction which is formally exact. In Sec.\,III,
proper approximations are made and the time-local interaction is obtained.
In Sec.\,IV, we study the dipole-dipole interaction in a finite volume.
Finally, we draw summary in Sec.\,V. Some calculation details are
presented in the Appendices.

\section{Retarded interaction between two magnetic dipoles }

We first consider there are two magnetic dipoles fixed in the EM field,
and the total Hamiltonian is $\hat{{\cal H}}=\hat{H}_{1}+\hat{H}_{2}+\hat{H}_{\text{\textsc{em}}}+\hat{H}_{\mathrm{int}}$.
Here $\hat{H}_{1,2}$ are self-Hamiltonians of the two dipoles, which
are modeled as two-level systems ($|\mathsf{g}_{i}\rangle$, $|\mathsf{e}_{i}\rangle$),
and $\hat{H}_{i}=\hbar\Omega_{i}|\mathsf{e}_{i}\rangle\langle\mathsf{e}_{i}|$
for $i=1,2$. $\hat{H}_{\text{\textsc{em}}}$ represents the Hamiltonian
of the EM field. And $\hat{H}_{\mathrm{int}}$ is the interaction
Hamiltonian between the magnetic dipoles and the field (Appendix \ref{sec:The-dipole})
\cite{weinberg_lectures_2012}
\begin{equation}
\hat{H}_{\mathrm{int}}=-\hat{\boldsymbol{\mathfrak{m}}}_{1}\cdot\hat{\mathbf{B}}(\mathbf{x}_{1})-\hat{\boldsymbol{\mathfrak{m}}}_{2}\cdot\hat{\mathbf{B}}(\mathbf{x}_{2}),
\end{equation}
where $\hat{\boldsymbol{\mathfrak{m}}}_{i}$ is the magnetic dipole
operator, and $\mathbf{x}_{i}$ is the position of dipole-$i$. The
magnetic field operator, $\hat{\mathbf{B}}(\mathbf{x})=\nabla\times\hat{\mathbf{A}}(\mathbf{x})$,
reads as
\begin{equation}
\hat{\mathbf{B}}(\mathbf{x})=\sum_{\mathbf{k},\sigma}i\hat{\mathrm{e}}_{\mathbf{k}\check{\sigma}}\overline{Z}_{k}(\hat{a}_{\mathbf{k}\sigma}e^{i\mathbf{k}\cdot\mathbf{x}}-\hat{a}_{\mathbf{k}\sigma}^{\dagger}e^{-i\mathbf{k}\cdot\mathbf{x}}),\label{eq:B(r)}
\end{equation}
 where $\overline{Z}_{k}:=\sqrt{\mu_{0}\hbar\omega_{k}/2V}$, and
$\hat{\mathrm{e}}_{\mathbf{k}\check{\sigma}}:=\hat{\mathrm{e}}_{\mathbf{k}}\times\hat{\mathrm{e}}_{\mathbf{k}\sigma}$.
The index $\check{\sigma}$ means the polarization direction orthogonal
to $\hat{\mathrm{e}}_{\mathbf{k}\sigma}$. 

The magnetic dipole operator should be treated more carefully. Generally,
the dipole operator can be written as 
\[
\hat{\boldsymbol{\mathfrak{m}}}=\big(\vec{m}_{\mathsf{ee}}|\mathsf{e}\rangle\langle\mathsf{e}|+\vec{m}_{\mathsf{gg}}|\mathsf{g}\rangle\langle\mathsf{g}|\big)+\big(\vec{m}_{\mathsf{eg}}|\mathsf{e}\rangle\langle\mathsf{g}|+\mathbf{h.c.}\big),
\]
 where $\vec{m}_{xy}:=\langle x|\hat{\boldsymbol{\mathfrak{m}}}|y\rangle$
for $x,y=\mathsf{e},\mathsf{g}$. The diagonal part should be regarded
as the permanent dipole, since it means the expectation value of the
dipole moment on each level;\emph{ }the off-diagonal part is the transition
dipole, which is widely discussed in radiation problems.

Therefore, for the above two dipoles, we denote $\hat{\boldsymbol{\mathfrak{m}}}_{i}=(\hat{\boldsymbol{\mathfrak{m}}}_{i}^{\mathsf{e}}+\hat{\boldsymbol{\mathfrak{m}}}_{i}^{\mathsf{g}})+\hat{\boldsymbol{\mathfrak{m}}}_{i}^{\text{\textsc{t}}}$,
where 
\begin{align}
\hat{\boldsymbol{\mathfrak{m}}}_{i}^{\text{\textsc{t}}} & =\vec{m}_{i}^{\text{\textsc{t}}}(|\mathsf{e}_{i}\rangle\langle\mathsf{g}_{i}|+|\mathsf{e}_{i}\rangle\langle\mathsf{g}_{i}|):=\vec{m}_{i}^{\text{\textsc{t}}}\,\hat{\tau}_{i}^{\text{\textsc{t}}},\nonumber \\
\hat{\boldsymbol{\mathfrak{m}}}_{i}^{\mathsf{e}} & =\vec{m}_{i}^{\mathsf{e}}|\mathsf{e}_{i}\rangle\langle\mathsf{e}_{i}|:=\vec{m}_{i}^{\mathsf{e}}\,\hat{\tau}_{i}^{\mathsf{e}},\label{eq:M}\\
\hat{\boldsymbol{\mathfrak{m}}}_{i}^{\mathsf{g}} & =\vec{m}_{i}^{\mathsf{g}}|\mathsf{g}_{i}\rangle\langle\mathsf{g}_{i}|:=\vec{m}_{i}^{\mathsf{g}}\,\hat{\tau}_{i}^{\mathsf{g}}.\nonumber 
\end{align}
Here $\hat{\tau}_{i}^{\text{\textsc{t}}}:=|\mathsf{e}_{i}\rangle\langle\mathsf{g}_{i}|+|\mathsf{e}_{i}\rangle\langle\mathsf{g}_{i}|$
and $\hat{\tau}_{i}^{\mathsf{e}(\mathsf{g})}:=|\mathsf{e}_{i}(\mathsf{g}_{i})\rangle\langle\mathsf{e}_{i}(\mathsf{g}_{i})|$
are unitless operators. A certain phase is chosen to make sure $\vec{m}_{i}^{\text{\textsc{t}}}=\langle\mathsf{e}_{i}|\hat{\boldsymbol{\mathfrak{m}}}_{i}^{\text{\textsc{t}}}|\mathsf{g}_{i}\rangle$
is real. $\hat{\boldsymbol{\mathfrak{m}}}_{i}^{\mathsf{e},\mathsf{g}}$
are the permanent dipole operators, where $\vec{m}_{i}^{\mathsf{e}}:=\langle\mathsf{e}_{i}|\hat{\boldsymbol{\mathfrak{m}}}|\mathsf{e}_{i}\rangle$
and $\vec{m}_{i}^{\mathsf{g}}:=\langle\mathsf{g}_{i}|\hat{\boldsymbol{\mathfrak{m}}}|\mathsf{g}_{i}\rangle$
are the permanent dipole moments on $|\mathsf{e}_{i}\rangle$, $|\mathsf{g}_{i}\rangle$
correspondingly, and they do not have to be equal to each other. Later
we will see that the permanent and transition dipole operators indeed
show quite different behaviors in dynamics, as well as the field induced
interaction.

With these notation, the interaction Hamiltonian is rewritten as $\hat{H}_{\mathrm{int}}=\sum_{i,\mu}\hat{\tau}_{i}^{\mu}\hat{B}_{i}^{\mu}$
for $i=1,2$ and $\mu=\text{\textsc{t}},\mathsf{e},\mathsf{g}$, where
\begin{align}
\hat{B}_{i}^{\mu} & =\sum_{\mathbf{k}\sigma}g_{i,\mathbf{k}\sigma}^{\mu}\hat{a}_{\mathbf{k}\sigma}+(g_{i,\mathbf{k}\sigma}^{\mu})^{*}\hat{a}_{\mathbf{k}\sigma}^{\dagger},\nonumber \\
g_{i,\mathbf{k}\sigma}^{\mu} & =-i(\vec{m}_{i}^{\mu}\cdot\hat{\mathrm{e}}_{\mathbf{k}\check{\sigma}})\overline{Z}_{k}e^{i\mathbf{k}\cdot\mathbf{x}_{i}}.
\end{align}
The coefficients $g_{i,\mathbf{k}\sigma}^{\mu}$ enclose contributions
from the EM field, the dipole moments ($\vec{m}_{i}^{\mu}$), and
the positions (\textbf{$\mathbf{x}_{i}$}).

Now we derive the dipole-dipole interaction induced by field. First,
considering only dipole-1 is placed in the field, due to the interaction
with dipole-1, the field dynamics is given by the Heisenberg equation
as 
\begin{align}
\partial_{t}\hat{a}_{\mathbf{k}\sigma} & =-i\omega_{\mathbf{k}}\hat{a}_{\mathbf{k}\sigma}-\sum_{\mu}^{\text{\textsc{t}},\mathsf{e},\mathsf{g}}\frac{i}{\hbar}(g_{1,\mathbf{k}\sigma}^{\mu})^{*}\,\hat{\tau}_{1}^{\mu},\\
\hat{a}_{\mathbf{k}\sigma}(t) & =\hat{a}_{\mathbf{k}\sigma}(0)e^{\text{--}i\omega_{\mathbf{k}}t}-\sum_{\mu}^{\text{\textsc{t}},\mathsf{e},\mathsf{g}}\frac{i(g_{1,\mathbf{k}\sigma}^{\mu})^{*}}{\hbar}\int_{0}^{t}ds\,e^{\text{--}i\omega_{\mathbf{k}}(t\text{--}s)}\hat{\tau}_{1}^{\mu}(s).\nonumber 
\end{align}
The first term in $\hat{a}_{\mathbf{k}\sigma}(t)$ comes from the
free evolution of the EM field, and the second term comes from the
interaction with dipole-1.

Then we put this $\hat{a}_{\mathbf{k}\sigma}(t)$ into the field operator
Eq.\,(\ref{eq:B(r)}), and the magnetic field can be arranged as
$\hat{\mathbf{B}}(\mathbf{x},t)=\hat{\mathbf{B}}_{0}(\mathbf{x},t)+\hat{\mathbf{B}}_{1}(\mathbf{x},t)$,
where 
\begin{align}
\hat{\mathbf{B}}_{0} & =\sum_{\mathbf{k}\sigma}i\hat{\mathrm{e}}_{\mathbf{k}\check{\sigma}}\overline{Z}_{k}\left[\hat{a}_{\mathbf{k}\sigma}(0)e^{i\mathbf{k}\cdot\mathbf{x}-i\omega_{k}t}-\mathbf{h.c.}\right],\\
\hat{\mathbf{B}}_{1} & =\sum_{\mathbf{k}\sigma,\mu}\frac{\hat{\mathrm{e}}_{\mathbf{k}\check{\sigma}}\overline{Z}_{k}e^{i\mathbf{k}\cdot\mathbf{x}}}{\hbar}(g_{1,\mathbf{k}\sigma}^{\mu})^{*}\int_{0}^{t}ds\,e^{-i\omega_{\mathbf{k}}(t-s)}\hat{\tau}_{1}^{\mu}(s)+\mathbf{h.c.}\nonumber 
\end{align}
are the vacuum field and the dipole field correspondingly.

Now we consider dipole-2 is put into the field, and interacts with
the EM field via $-\hat{\boldsymbol{\mathfrak{m}}}_{2}\cdot\hat{\mathbf{B}}(\mathbf{x}_{2})$.
The dipole-dipole interaction is induced by the dipole field $\hat{\mathbf{B}}_{1}(\mathbf{x},t)$,
which gives (denoting $s':=t-s$) \begin{widetext}
\begin{align}
\hat{H}_{12} & =-\hat{\boldsymbol{\mathfrak{m}}}_{2}\cdot\hat{\mathbf{B}}_{1}(\mathbf{x}_{2})=\sum_{\mathbf{k}\sigma,\mu\nu}-\frac{i}{\hbar}(g_{1,\mathbf{k}\sigma}^{\mu})^{*}g_{2,\mathbf{k}\sigma}^{\nu}\int_{0}^{t}ds\,e^{-i\omega_{\mathbf{k}}(t-s)}\hat{\tau}_{1}^{\mu}(s)\hat{\tau}_{2}^{\nu}(t)+\mathbf{h.c.}\nonumber \\
 & =\sum_{\mu\nu}^{\text{\textsc{p,t}}}-\frac{i}{\hbar}\int_{0}^{t}ds'\left(\int_{0}^{\infty}\frac{d\omega}{2\pi}\,\big[J_{12}^{\mu\nu}(\omega)e^{-i\omega s'}-J_{21}^{\mu\nu}(\omega)e^{i\omega s'}\big]\right)\hat{\tau}_{1}^{\mu}(t-s')\hat{\tau}_{2}^{\nu}(t).\label{eq:retard}
\end{align}
\end{widetext} Here $J_{ij}^{\mu\nu}(\omega):=2\pi\sum_{\mathbf{k}\sigma}(g_{i,\mathbf{k}\sigma}^{\mu})^{*}g_{j,\mathbf{k}\sigma}^{\nu}\delta(\omega-\omega_{\mathbf{k}})$
is the coupling spectral density ($i,j=1,2$, and $\mu,\nu=\text{\textsc{t}},\mathsf{e},\mathsf{g}$)
\cite{breuer_theory_2002,li_steady_2015}, which is adopted to convert
the summation into an integral.

We should also consider the reverse procedure, i.e., first put dipole-2
in the field, then consider the interaction between dipole-1 and the
field generated by dipole-2. That gives a Hamiltonian $\hat{H}_{21}$,
and the complete dipole-dipole interaction should be $(\hat{H}_{12}+\hat{H}_{21})/2$.

Up to now, $\hat{H}_{12}$ {[}Eq.\,(\ref{eq:retard}){]} is an exact
result, and quite naturally, it has a retarded form, which indicates
the interaction between the two dipoles is not instantaneous. This
Hamiltonian contains interaction of both the permanent and transition
dipoles, and the transition frequencies do not have to be resonant
with each other. 

\section{Time-local interaction}

Here we further adopt several approximations to get a time-local interaction.
Since Eq.\,(\ref{eq:retard}) is already in the 2nd order of the
interaction strength $g_{i,\mathbf{k}\sigma}^{\mu}$, approximately
we only keep the 0-th order of $\hat{\tau}_{i}^{\text{\textsc{t}},\mathsf{e},\mathsf{g}}(t)$
which is governed by $\hat{H}_{i}=\hbar\Omega_{i}|\mathsf{e}_{i}\rangle\langle\mathsf{e}_{i}|$,
and that is (considering the resonance case $\Omega_{1}=\Omega_{2}:=\Omega$)
\cite{lehmberg_radiation_1970}
\begin{equation}
\hat{\tau}_{i}^{\mathsf{e}(\mathsf{g})}(t)\simeq\hat{\tau}_{i}^{\mathsf{e}(\mathsf{g})},\qquad\hat{\tau}_{i}^{\text{\textsc{t}}}(t)\simeq\hat{\tau}_{i}^{-}e^{-i\Omega t}+\hat{\tau}_{i}^{+}e^{i\Omega t},
\end{equation}
where $\hat{\tau}_{i}^{-}:=|\mathsf{g}_{i}\rangle\langle\mathsf{e}_{i}|$
and $\hat{\tau}_{i}^{+}:=|\mathsf{e}_{i}\rangle\langle\mathsf{g}_{i}|$. 

Clearly, the permanent and transition dipoles show quite different
behaviors in dynamics. The transition dipole contains a rotation with
frequency $\Omega$, but the permanent dipoles $\hat{\tau}_{i}^{\mathsf{e},\mathsf{g}}(t)$
are ``static'' and independent of time, since they only contains
diagonal elements. Or we can also regard them as rotating with zero
frequency. Below we will see such a distinction in their dynamics
also influences the behavior when they exchange interactions through
the field.

We apply RWA to the term $\hat{\tau}_{1}^{\mu}(t-s')\hat{\tau}_{2}^{\nu}(t)$
\cite{breuer_theory_2002,li_long-term_2014,li_steady_2015}, and omit
the fast-oscillating terms with coefficients $e^{\pm i\Omega t}$
or $e^{\pm2i\Omega t}$, then the remaining terms are $\hat{\tau}_{2}^{+}\hat{\tau}_{1}^{-}e^{i\Omega s'}$,
$\hat{\tau}_{2}^{-}\hat{\tau}_{1}^{+}e^{-i\Omega s'}$ and $\hat{\tau}_{1}^{x}\hat{\tau}_{2}^{y}$
($x,y=\mathsf{e},\mathsf{g}$). The first two terms describe the transition
dipole interaction, and the third one describes the permanent dipole
interaction. Again we see they contains the oscillating frequency
of $\Omega$ and 0 respectively.

\vspace{1em}\noindent\textbf{\emph{ Transition dipole:}} We first
look at the interaction between transition dipoles. Put $\hat{\tau}_{2}^{+}\hat{\tau}_{1}^{-}e^{i\Omega s'}$,
$\hat{\tau}_{2}^{-}\hat{\tau}_{1}^{+}e^{-i\Omega s'}$ into Eq.\,(\ref{eq:retard}),
and that gives 
\begin{align}
\hat{H}_{12}^{\text{\textsc{t}}} & =-\frac{i}{\hbar}\int_{0}^{t}ds'\int_{0}^{\infty}\frac{d\omega}{2\pi}\,\big[J_{12}^{\text{\textsc{t}}\text{\textsc{t}}}(\omega)e^{-i\omega s'}-J_{21}^{\text{\textsc{t}}\text{\textsc{t}}}(\omega)e^{i\omega s'}\big]\nonumber \\
 & \qquad\times(\hat{\tau}_{2}^{+}\hat{\tau}_{1}^{-}e^{i\Omega s'}+\hat{\tau}_{2}^{-}\hat{\tau}_{1}^{+}e^{-i\Omega s'}).
\end{align}
Usually the dipole-dipole interaction is established after very short
time, thus the upper limit of the above time integral can be extended
to $\infty$ (Markovian approximation) \cite{gardiner_quantum_2004,li_production_2017}.
After the time integration, we obtain $\hat{H}_{12}^{\text{\textsc{t}}}=\xi^{\text{\textsc{t}}}(\hat{\tau}_{2}^{+}\hat{\tau}_{1}^{-}+\hat{\tau}_{2}^{-}\hat{\tau}_{1}^{+})$
\protect\footnote{ Here we need the formula $\int_{0}^{\infty}dt\,e^{i(\omega-\Omega)t}=\pi\delta(\omega-\Omega)+i\mathbf{P}\frac{1}{\omega-\Omega}$.
}, where the interaction strength $\xi^{\text{\textsc{t}}}$ is obtained
by substituting the coupling spectral density for the EM field into
the above integral, and that gives {[}here we have $J_{12}^{\text{\textsc{tt}}}(-\omega)=-J_{12}^{\text{\textsc{tt}}}(\omega)$,
and $J_{12}^{\text{\textsc{tt}}}(\omega)=J_{21}^{\text{\textsc{tt}}}(\omega)$,
see Appendix \ref{sec:Coupling-spectral-density}{]} 
\begin{align}
 & \xi^{\text{\textsc{t}}}=-\int_{-\infty}^{\infty}\frac{d\omega}{2\pi\hbar}\,\frac{J_{12}^{\text{\textsc{tt}}}(\omega)}{\omega-\Omega}\label{eq:T-free}\\
= & \frac{\mu_{0}}{4\pi r^{3}}\Big\{-\big[\vec{m}_{1}^{\text{\textsc{t}}}\cdot\vec{m}_{2}^{\text{\textsc{t}}}-(\vec{m}_{1}^{\text{\textsc{t}}}\cdot\hat{\mathrm{e}}_{\mathbf{r}})(\vec{m}_{2}^{\text{\textsc{t}}}\cdot\hat{\mathrm{e}}_{\mathbf{r}})\big]x_{\Omega}^{2}\cos x_{\Omega}\nonumber \\
+ & \big[\vec{m}_{1}^{\text{\textsc{t}}}\cdot\vec{m}_{2}^{\text{\textsc{t}}}-3(\vec{m}_{1}^{\text{\textsc{t}}}\cdot\hat{\mathrm{e}}_{\mathbf{r}})(\vec{m}_{2}^{\text{\textsc{t}}}\cdot\hat{\mathrm{e}}_{\mathbf{r}})\big](\cos x_{\Omega}-x_{\Omega}\sin x_{\Omega})\Big\}.\nonumber 
\end{align}
Here we denote $x_{\Omega}:=2\pi r/\lambda$, $r:=|\mathbf{x}_{2}-\mathbf{x}_{1}|$
is the distance between the two dipoles, and $\lambda$ is the wavelength
of the transition frequency $\Omega$. In the short-distance limit
$r/\lambda\rightarrow0$, this interaction strength $\xi^{\text{\textsc{t}}}$
returns to the same form with the classical result {[}Eq.\,(\ref{eq:classical}){]}.
This is the same with the situation in previous studies about the
field induced dipole-dipole interaction where two resonant electric
transition dipoles were concerned \cite{agarwal_quantum_1974,ficek_quantum_1987,ficek_quantum_2005,lehmberg_radiation_1970}.

\vspace{1em}\noindent\textbf{\emph{ Permanent dipole:}} Now we consider
the remaining terms $\hat{\tau}_{1}^{x}\hat{\tau}_{2}^{y}$ ($x,y=\mathsf{e},\mathsf{g}$)
of RWA, which indicate the permanent dipole interactions. Following
the same approach as above, the interaction Hamiltonian gives $\hat{H}_{12}^{xy}=\xi^{xy}\hat{\tau}_{1}^{x}\hat{\tau}_{2}^{y}$,
where the interaction strength $\xi^{xy}$ is ($x,y=\mathsf{e},\mathsf{g}$)
\begin{align}
\xi^{xy}= & -\int_{-\infty}^{\infty}\frac{d\omega}{2\pi\hbar}\,\frac{J_{12}^{xy}(\omega)}{\omega}\nonumber \\
= & \frac{\mu_{0}}{4\pi r^{3}}\big[\vec{m}_{1}^{x}\cdot\vec{m}_{2}^{y}-3(\vec{m}_{1}^{x}\cdot\hat{\mathrm{e}}_{\mathbf{r}})(\vec{m}_{2}^{y}\cdot\hat{\mathrm{e}}_{\mathbf{r}})\big].\label{eq:P-free}
\end{align}
 This integral can be also regarded as setting $\Omega=0$ in Eq.\,(\ref{eq:T-free}),
since the permanent dipoles are ``static'' and do not oscillate,
as we mentioned before (the rotating frequency is 0).

Therefore, the interaction between the two permanent dipoles is 
\begin{equation}
\hat{H}_{12}^{xy}=\frac{\mu_{0}}{4\pi r^{3}}\big[\hat{\boldsymbol{\mathfrak{m}}}_{1}^{x}\cdot\hat{\boldsymbol{\mathfrak{m}}}_{2}^{y}-3(\hat{\boldsymbol{\mathfrak{m}}}_{1}^{x}\cdot\hat{\mathrm{e}}_{\mathbf{r}})(\hat{\boldsymbol{\mathfrak{m}}}_{2}^{y}\cdot\hat{\mathrm{e}}_{\mathbf{r}})\big],
\end{equation}
for $x,y=\mathsf{e},\mathsf{g}$. Remember $\hat{\boldsymbol{\mathfrak{m}}}_{i}^{\mathsf{e}}=\vec{m}_{i}^{\mathsf{e}}|\mathsf{e}_{i}\rangle\langle\mathsf{e}_{i}|$
and $\hat{\boldsymbol{\mathfrak{m}}}_{i}^{\mathsf{g}}=\vec{m}_{i}^{\mathsf{g}}|\mathsf{g}_{i}\rangle\langle\mathsf{g}_{i}|$
are the dipole operators for $|\mathsf{e}_{i}\rangle$ and $|\mathsf{g}_{i}\rangle$
respectively, and the values $\vec{m}_{i}^{\mathsf{e}}$ and $\vec{m}_{i}^{\mathsf{g}}$
do not have to equal to each other. $\hat{H}_{12}^{xy}$ describes
the permanent dipole-dipole interaction when the two dipoles stay
in states $|x\rangle$ and $|y\rangle$ respectively, and it has exactly
the same form with the classical magnetic dipole-dipole interaction
{[}Eq.\,(\ref{eq:classical}){]}. 

Notice that the above derivation process also implies that this interaction
does not relies on the state of the EM field, e.g., whether it is
in a thermal state or squeezed state. The generalization to multilevel
systems is straightforward. This result indicates that the diagonal
part of the dipole operator corresponds to classical physics, and
the off-diagonal part contains quantum corrections.

\section{Interaction in a finite periodic box}

We have shown that the interaction between two remote dipoles can
be derived through their interaction with the EM field. Especially,
the interaction between permanent dipoles exactly returns to the classical
result. Then an intriguing question arises: if the property of the
EM field is changed, whether the dipole-dipole interaction can be
changed. 

Here we consider that the two dipoles are confined in a box with a
finite volume $V=L^{3}$, and the modes of the EM field are highly
discrete. This can be realized by 3D metal cavity in experiments \cite{rigetti_superconducting_2012,paik_observation_2011}.
For simplicity, here we consider a box geometry with periodic boundary
condition, thus, the coupling strengths $g_{i,\mathbf{k}\sigma}^{\mu}$
are the same with the above calculations.

The derivation for the dipole-dipole interaction follows the same
way as the above procedure, but we should pause at the first line
of Eq.\,(\ref{eq:retard}), where the summation cannot be turned
into integral now. After RWA and Markovian approximation as before,
we still obtain permanent/transition dipole-dipole interactions as
$\hat{H}_{12}^{xy}=\xi^{xy}\hat{\tau}_{1}^{x}\hat{\tau}_{2}^{y}$
and $\hat{H}_{12}^{\text{\textsc{t}}}=\xi^{\text{\textsc{t}}}(\hat{\tau}_{2}^{+}\hat{\tau}_{1}^{-}+\hat{\tau}_{2}^{-}\hat{\tau}_{1}^{+})$,
except the coupling strengths should be recalculated.

\begin{figure}
\includegraphics[width=0.8\columnwidth]{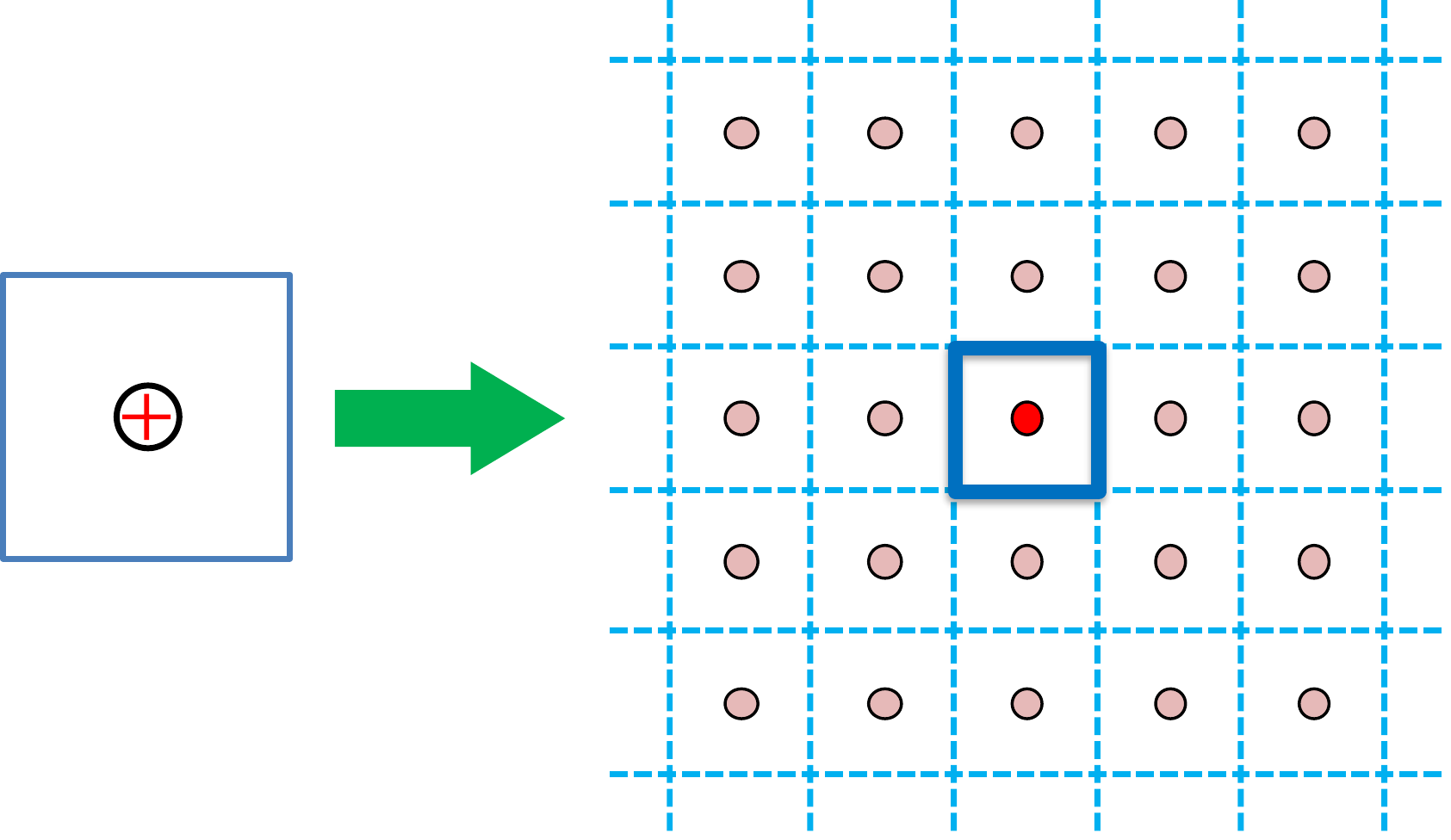}

\caption{(Color online) Demonstration for the boundary condition of Eq.\,(\ref{eq:potential}).
The influence of the periodic boundary condition can be equivalently
replaced by the point ``charge'' lattice in free space, which is
similar like the method of image charges.}

\label{fig-boundary}
\end{figure}

\vspace{1em}\noindent\textbf{\emph{ Permanent dipole:}} We first
look at the interaction of permanent dipoles $\hat{H}_{12}^{xy}=\xi^{xy}\hat{\tau}_{1}^{x}\hat{\tau}_{2}^{y}$
($x,y=\mathsf{e},\mathsf{g}$). After the time integration as above,
the interaction strength $\xi^{xy}$ is given by ($\mathbf{r}:=\mathbf{x}_{2}-\mathbf{x}_{1}$)
\begin{align}
\xi^{xy} & =\sum_{\mathbf{k},\sigma}-\frac{(g_{1,\mathbf{k}\sigma}^{x})^{*}g_{2,\mathbf{k}\sigma}^{y}}{\hbar\omega_{\mathbf{k}}}+\mathbf{h.c.}\label{eq:coupling-p}\\
= & \sum_{\mathbf{k}\neq0}-\frac{\mu_{0}}{2V}\Big[\vec{m}_{1}^{x}\cdot\vec{m}_{2}^{y}+(\vec{m}_{1}^{x}\cdot\nabla_{\mathbf{r}})(\vec{m}_{2}^{y}\cdot\nabla_{\mathbf{r}})\frac{1}{\mathbf{k}^{2}}\Big]e^{i\mathbf{k}\cdot\mathbf{r}}+\mathbf{h.c.}\nonumber 
\end{align}
Utilizing the normalization relation of $e^{i\mathbf{k}\cdot\mathbf{r}}$
inside the periodic box \footnote{\protect Notice that $\{\varphi_{\mathbf{k}}|\varphi_{\mathbf{k}}(\mathbf{r})=e^{i\mathbf{k}\cdot\mathbf{r}}/\sqrt{V}\}$
is an orthonormal set, and the wavefunction $\delta^{(3)}(\mathbf{r})$
can be expanded as $\delta^{(3)}(\mathbf{r})=\sum_{\mathbf{k}}\alpha_{\mathbf{k}}\varphi_{\mathbf{k}}(\mathbf{r)}$
with $\alpha_{\mathbf{k}}=1/\sqrt{V}.$}, the first term in the above
summation leads to 
\begin{equation}
-\mu_{0}\vec{m}_{1}^{x}\cdot\vec{m}_{2}^{y}\big[\sum_{\mathbf{k}\neq0}\frac{e^{i\mathbf{k}\cdot\mathbf{r}}}{V}\big]=\frac{\mu_{0}}{V}\vec{m}_{1}^{x}\cdot\vec{m}_{2}^{y}\big[1-V\delta^{(3)}(\mathbf{r})\big].
\end{equation}

The second summation term can be calculated by $-\mu_{0}(\vec{m}_{1}^{x}\cdot\nabla_{\mathbf{r}})(\vec{m}_{2}^{y}\cdot\nabla_{\mathbf{r}})\chi_{\text{\textsc{p}}}(\mathbf{r})$,
where 
\begin{equation}
\chi_{\text{\textsc{p}}}(\mathbf{r})=\sum_{\mathbf{k}\neq0}\frac{1}{\mathbf{k}^{2}}\frac{e^{i\mathbf{k}\cdot\mathbf{r}}}{V}
\end{equation}
is a generation function. For a finite volume, the field modes are
discrete, still the summation in $\chi_{\text{\textsc{p}}}(\mathbf{r})$
cannot be turned into integral. Notice that the generation function
$\chi_{\text{\textsc{p}}}(\mathbf{r})$ is a periodic function $\chi_{\text{\textsc{p}}}(\mathbf{r})=\chi_{\text{\textsc{p}}}(\mathbf{r}+\mathbf{r}_{\mathbf{n}})$,
where $\mathbf{r}_{\mathbf{n}}:=L(n_{x},n_{y},n_{z})$ is a periodicity
vector ($n_{i}$ are integers), and it satisfies the following differential
equation
\begin{equation}
\nabla^{2}\chi_{\text{\textsc{p}}}(\mathbf{r})=-\sum_{\mathbf{k}\neq0}\frac{e^{i\mathbf{k}\cdot\mathbf{r}}}{V}=-\sum_{\mathbf{n}}\delta^{(3)}(\mathbf{r}-\mathbf{r}_{\mathbf{n}})+\frac{1}{V}.\label{eq:potential}
\end{equation}

Here $\chi_{\text{\textsc{p}}}(\mathbf{r})$ can be regarded as an
``electric potential'' in free space, with $-\sum\delta^{(3)}(\mathbf{r}-\mathbf{r}_{n})$
as negative ``charges'' at $\mathbf{r}_{\mathbf{n}}$, and $1/V$
as a homogenous positive ``background'' (Fig.\,\ref{fig-boundary}).
Thus, the solution of the above equation is
\begin{equation}
\chi_{\text{\textsc{p}}}(\mathbf{r})=\sum_{\mathbf{n}}\frac{1}{4\pi}\frac{-1}{\left|\mathbf{r}-\mathbf{r}_{\mathbf{n}}\right|}+\frac{\mathbf{r}^{2}}{2V}.
\end{equation}
Therefore, the permanent dipole interaction strength is
\begin{align}
\xi^{xy} & =\frac{\mu_{0}\,\vec{m}_{1}^{x}\cdot\vec{m}_{2}^{y}}{V}-\mu_{0}(\vec{m}_{1}^{x}\cdot\nabla_{\mathbf{r}})(\vec{m}_{2}^{y}\cdot\nabla_{\mathbf{r}})\chi_{\text{\textsc{p}}}(\mathbf{r})\label{eq:P-box}\\
 & =\frac{\mu_{0}}{4\pi}\sum_{\mathbf{n}}\frac{\vec{m}_{1}^{x}\cdot\vec{m}_{2}^{y}-3(\vec{m}_{1}^{x}\cdot\hat{\mathrm{e}}_{\delta\mathbf{r}_{\mathbf{n}}})(\vec{m}_{2}^{y}\cdot\hat{\mathrm{e}}_{\delta\mathbf{r}_{\mathbf{n}}})}{|\Delta\mathbf{r}_{\mathbf{n}}|^{3}},\nonumber 
\end{align}
where $\Delta\mathbf{r}_{\mathbf{n}}:=\mathbf{r}-\mathbf{r}_{\mathbf{n}}$,
and $\hat{\mathrm{e}}_{\delta\mathbf{r}_{\mathbf{n}}}:=\Delta\mathbf{r}_{\mathbf{n}}/|\Delta\mathbf{r}_{\mathbf{n}}|$.

\begin{figure}
\includegraphics[width=0.75\columnwidth]{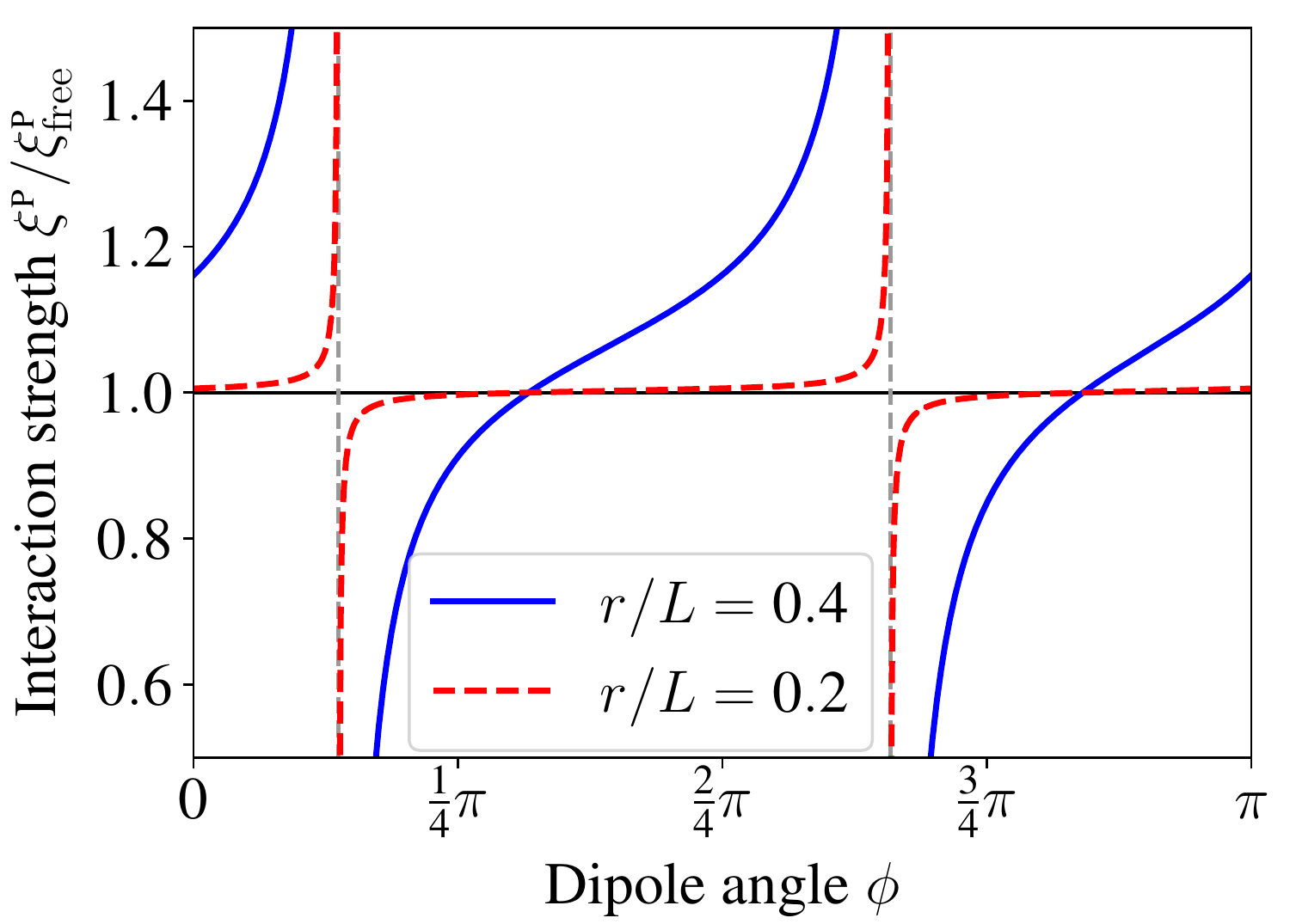}

\caption{(Color online) Numerical estimation for the dipole-dipole interaction
strength in a finite volume {[}Eq.\,(\ref{eq:P-box}){]} comparing
with the free space result. Here as an example, we choose $\hat{\mathrm{e}}_{\mathbf{r}}=(1,2,3)/\sqrt{14}$,
and set the two dipoles parallel to each other in the direction $\hat{\mathrm{e}}_{1,2}=(\cos\phi,0,\sin\phi)$.
When $r/L$ is large, the correction from the boundary effect is significant.
It turns out Eq.\,(\ref{eq:P-box}) converges quite fast, and only
very few ``image'' terms is needed ($|n_{x,y,z}|\lesssim2$) to
get a precise enough result, which means only the nearest ``image''
dipoles are important.}

\label{fig-correction}
\end{figure}

The above result $\xi^{xy}$ only depends on the relative distance
$\mathbf{r}=\mathbf{x}_{2}-\mathbf{x}_{1}$, but does not depend the
absolute positions $\mathbf{x}_{1,2}$. This is because the box with
periodic boundary condition is translationally symmetric, thus every
point can be regarded as the box center. Without generality, we set
the position of dipole-1 ($\mathbf{x}_{1}$) as the origin, then the
position of dipole-2 is $\mathbf{x}_{2}=\mathbf{r}$.

Notice that, the 0-th term {[}$\mathbf{n}=(0,0,0)${]} in the above
summation is exactly the same with the free space result Eq.\,(\ref{eq:P-free}).
The other summation terms can be regarded as contributions from ``image''
dipoles of dipole-2 at the positions $\mathbf{r}-\mathbf{r}_{\mathbf{n}}$
reflecting the boundary effect, and they are of order $\sim V^{-1}$.
Thus, when $V\rightarrow\infty$, this result returns to the previous
free space case.

Fig.\,\ref{fig-correction} shows a numerical estimation for this
permanent dipole interaction strength in a finite volume {[}Eq.\,(\ref{eq:P-box}){]}
comparing with the free space result {[}Eq.\,(\ref{eq:P-free}){]}.
In a finite volume, the dipole-dipole interaction strength can be
either enhanced ($\xi^{xy}/\xi_{\mathrm{free}}^{xy}>1$) or decreased
($\xi^{xy}/\xi_{\mathrm{free}}^{xy}<1$), depending on the dipole
orientations. There are two diverging points shown in the figure (dashed
gray line), this is because around these dipole orientations, $\vec{m}_{1}^{x}\cdot\vec{m}_{2}^{y}-3(\vec{m}_{1}^{x}\cdot\hat{\mathrm{e}}_{\mathbf{r}})(\vec{m}_{2}^{y}\cdot\hat{\mathrm{e}}_{\mathbf{r}})\simeq0$,
thus the free space result approaches zero, which makes $\xi^{xy}/\xi_{\mathrm{free}}^{xy}$
diverge.

If both the two dipoles are placed far away from the boundary, we
have $r/L\ll1$, thus only the 0-th term is important, and that means
the dipole interaction is almost the same with free space case. On
the other hand, if they are close to the boundary, effectively they
get close to the ``image'' dipoles, thus the correction from the
boundary effect becomes significant. This is also quite similar with
the situation in classical electrodynamics.

\vspace{1em}\noindent\textbf{\emph{ Transition dipole:}} Now we consider
the interaction between two transition dipoles. The interaction strength
is
\begin{align}
\xi^{\text{\textsc{t}}} & =\sum_{\mathbf{k},\sigma}-\frac{(g_{1,\mathbf{k}\sigma}^{\text{\textsc{t}}})^{*}g_{2,\mathbf{k}\sigma}^{\text{\textsc{t}}}}{\hbar\omega_{\mathbf{k}}}[1+\frac{\Omega}{\omega_{\mathbf{k}}-\Omega}]+\mathbf{h.c.}\\
 & =\xi_{0}^{\text{\textsc{t}}}+\frac{\mu_{0}\,\vec{m}_{1}^{\text{\textsc{t}}}\cdot\vec{m}_{2}^{\text{\textsc{t}}}}{V}-\mu_{0}(\vec{m}_{1}^{\text{\textsc{t}}}\cdot\nabla_{\mathbf{r}})(\vec{m}_{2}^{\text{\textsc{t}}}\cdot\nabla_{\mathbf{r}})\chi_{\text{\textsc{t}}}(\mathbf{r}).\nonumber 
\end{align}
Here $\xi_{0}^{\text{\textsc{t}}}$ has the same form with the permanent
dipole interaction Eq.\,(\ref{eq:P-box}), except the dipole index
should be changed to ``$\text{\textsc{t}}$'', and $\chi_{\text{\textsc{t}}}$
is a generation function:
\begin{equation}
\chi_{\text{\textsc{t}}}(\mathbf{r}):=\sum_{\mathbf{k}\neq0}\frac{\Omega}{\omega_{\mathbf{k}}-\Omega}\cdot\frac{e^{i\mathbf{k}\cdot\mathbf{r}}}{\mathbf{k}^{2}}\cdot\frac{1}{V}.
\end{equation}
 Comparing with the generation function $\chi_{\text{\textsc{p}}}$
in the permanent dipoles case, here $\chi_{\text{\textsc{t}}}$ contains
a sharp envelop $\Omega/(\omega_{\mathbf{k}}-\Omega)$ in the summation.
Therefore, only the nearly resonant terms with $\omega_{\mathbf{k}}\simeq\Omega$
contribute significantly in the summation, and they could even dominate
over $\xi_{0}^{\text{\textsc{t}}}$. Thus, the interaction strength
can be also approximately recalculated by 
\begin{equation}
\xi^{\text{\textsc{t}}}=\sum_{\mathbf{k}}^{\omega_{\mathbf{k}}\simeq\Omega}-\frac{(g_{1,\mathbf{k}\sigma}^{\text{\textsc{t}}})^{*}g_{2,\mathbf{k}\sigma}^{\text{\textsc{t}}}}{\hbar(\omega_{\mathbf{k}}-\Omega)}+\mathbf{h.c.}
\end{equation}

Notice that this result also has a similar form with some previous
studies based on adiabatic elimination ($2|g|^{2}/\Delta$) \cite{lambert_cavity-mediated_2016}.
Thus, the correction mechanism to the transition dipole interaction
in a finite volume is quite similar with the Purcell effect \cite{purcell_resonance_1946,baranov_magnetic_2016}. 

\section{Summary}

In this paper, we derived the indirect interaction between two magnetic
dipoles induced by the quantized EM field for both free space and
finite volume case. A retarded interaction is obtained, and it reduces
to a time-local one after RWA and Markovian approximation.

Our result showed that the permanent and transition dipoles should
be treated separately. In free space, the interaction between the
permanent dipoles directly returns to the classical interaction; the
interaction between transition dipoles has the same form with the
previous studies on electric dipole interaction, and it does not return
to the classical result directly, yet returns in the short-distance
limit $r/\lambda\ll1$.

In a finite volume, both the permanent and transition dipole-dipole
interactions are changed, but by different mechanisms. For transition
dipoles, this changing mechanism is similar with the Purcell effect,
since only a few number of nearly resonant modes take effect in the
interaction mediation; for permanent dipoles, still all the field
modes take effect for the interaction mediation, but the correction
comes from the boundary effect: if the dipoles are placed close to
the boundary, the influence is strong, if they are both placed far
away from the boundary, their interaction does not change too much
from the free space case.

\emph{Acknowledgement} - This study is supported by Office of Naval
Research (Award No. N00014-16-1-3054) and Robert A. Welch Foundation
(Grant No. A-1261). S.-W. Li is very grateful for the helpful discussions
with Z. Yi at Texas A\&M University.

\appendix
\begin{widetext}

\section{The interaction between a magnetic dipole and the EM field \label{sec:The-dipole}}

Here we derive the interaction between a magnetic dipole and the EM
field. We start from the Hamiltonian of a single atom coupled with
the EM field 
\begin{align}
\hat{H}_{S} & =\frac{[\hat{\mathbf{p}}-e\hat{\mathbf{A}}(\hat{\mathbf{x}})]^{2}}{2m}+\varphi(\hat{\mathbf{x}}-\mathbf{x}_{0}),\nonumber \\
\hat{H}_{\text{\textsc{em}}} & =\int dV\,\frac{1}{2}[\epsilon_{0}\mathbf{E}^{2}+\frac{1}{\mu_{0}}\,\mathbf{B}^{2}]=\sum_{\mathbf{k},\sigma}\frac{1}{2}\hbar\omega_{\mathbf{k}}(\hat{a}_{\mathbf{k}\sigma}^{\dagger}\hat{a}_{\mathbf{k}\sigma}+\hat{a}_{\mathbf{k}\sigma}\hat{a}_{\mathbf{k}\sigma}^{\dagger}).
\end{align}
Here $\varphi(\mathbf{x})$ is the electric potential, and $\hat{\mathbf{A}}(\mathbf{x})$
is the field operator
\begin{equation}
\hat{\mathbf{A}}(\mathbf{x})=\sum_{\mathbf{k},\sigma}\hat{\mathrm{e}}_{\mathbf{k}\sigma}\overline{N}_{k}(\hat{a}_{\mathbf{k}\sigma}e^{i\mathbf{k}\cdot\mathbf{x}}+\hat{a}_{\mathbf{k}\sigma}^{\dagger}e^{-i\mathbf{k}\cdot\mathbf{x}}),\qquad\overline{N}_{k}:=\sqrt{\frac{\hbar}{2\epsilon_{0}\omega_{\mathbf{k}}V}}.
\end{equation}
We omit the $\hat{\mathbf{A}}^{2}$ term in $\hat{H}_{S}$, and expand
$\hat{\mathbf{A}}(\hat{\mathbf{x}})$ around the nucleus position
$\mathbf{x}_{0}$ by $e^{i\mathbf{k}\cdot\hat{\mathbf{x}}}\simeq e^{i\mathbf{k}\cdot\mathbf{x}_{0}}\big[1+i\mathbf{k}\cdot\hat{\mathbf{r}}+\dots\big]$,
where $\hat{\mathbf{r}}:=\hat{\mathbf{x}}-\mathbf{x}_{0}$. The zeroth
order just gives the interaction of dipole approximation, $\hat{H}_{\mathrm{int}}^{(0)}=-\frac{e}{m}\hat{\mathbf{p}}\cdot\hat{\mathbf{A}}(\mathbf{x}_{0})$
\cite{weinberg_lectures_2012}.

Now we consider the 1st order in the expansion, which gives
\begin{equation}
\hat{H}_{\mathrm{int}}^{(1)}=\sum_{\mathbf{k},\sigma}-\frac{e}{m}\cdot\overline{N}_{k}(\hat{\mathbf{p}}\cdot\hat{\mathrm{e}}_{\mathbf{k}\sigma})(i\mathbf{k}\cdot\hat{\mathbf{r}})\,(\hat{a}_{\mathbf{k}\sigma}e^{i\mathbf{k}\cdot\mathbf{x}_{0}}-\hat{a}_{\mathbf{k}\sigma}^{\dagger}e^{-i\mathbf{k}\cdot\mathbf{x}_{0}}).
\end{equation}
With the help of the relation (denoting $\hat{\mathbf{L}}:=\hat{\mathbf{r}}\times\hat{\mathbf{p}}$)
\begin{align}
(\hat{\mathbf{p}}\cdot\hat{\mathrm{e}}_{\mathbf{k}\sigma})(i\mathbf{k}\cdot\hat{\mathbf{r}}) & =(\hat{\mathbf{p}}\cdot i\mathbf{k})(\hat{\mathbf{r}}\cdot\hat{\mathrm{e}}_{\mathbf{k}\sigma})+(\hat{\mathbf{r}}\times\hat{\mathbf{p}})\cdot(i\mathbf{k}\times\hat{\mathrm{e}}_{\mathbf{k}\sigma})\nonumber \\
 & =\frac{i|\mathbf{k}|}{2}[(\hat{\mathbf{p}}\cdot\hat{\mathrm{e}}_{\mathbf{k}\sigma})(\hat{\mathbf{r}}\cdot\hat{\mathrm{e}}_{\mathbf{k}})+(\hat{\mathbf{p}}\cdot\hat{\mathrm{e}}_{\mathbf{k}})(\hat{\mathbf{r}}\cdot\hat{\mathrm{e}}_{\mathbf{k}\sigma})]+\frac{1}{2}\hat{\mathbf{L}}\cdot(i\mathbf{k}\times\hat{\mathrm{e}}_{\mathbf{k}\sigma}),
\end{align}
the above interaction Hamiltonian can be written into two terms $\hat{H}_{\mathrm{int}}^{(1)}=\hat{H}_{\mathrm{md}}+\hat{H}_{\mathrm{eq}}$,
where $\hat{H}_{\mathrm{md}}$ is the interaction between the magnetic
dipole and the EM field (denoting $\hat{\boldsymbol{\mathfrak{m}}}_{L}:=\frac{e}{2m}\hat{\mathbf{L}}$)
\begin{equation}
\hat{H}_{\mathrm{md}}=\sum_{\mathbf{k},\sigma}-\frac{e}{2m}\,\hat{\mathbf{L}}\cdot(i\mathbf{k})\times\hat{\mathrm{e}}_{\mathbf{k}\sigma}\,\overline{N}_{k}(\hat{a}_{\mathbf{k}\sigma}e^{i\mathbf{k}\cdot\mathbf{x}_{0}}-\hat{a}_{\mathbf{k}\sigma}^{\dagger}e^{-i\mathbf{k}\cdot\mathbf{x}_{0}})=-\frac{e}{2m}\hat{\mathbf{L}}\cdot\nabla_{\mathbf{x}_{0}}\times\hat{\mathbf{A}}(\mathbf{x}_{0})=-\hat{\boldsymbol{\mathfrak{m}}}_{L}\cdot\hat{\mathbf{B}}(\mathbf{x}_{0}),
\end{equation}
and $\hat{H}_{\mathrm{eq}}$ gives the electric quadrupole interaction
(denoting $\hat{p}_{i}=\hat{\mathbf{p}}\cdot\hat{\mathrm{e}}_{i}$,
$\hat{r}_{i}=\hat{\mathbf{r}}\cdot\hat{\mathrm{e}}_{i}$) \cite{weinberg_lectures_2012}
\begin{align}
\hat{H}_{\mathrm{eq}} & =\sum_{\mathbf{k},\sigma}-\frac{i|\mathbf{k}|e}{2m}\,\big(\hat{p}_{\mathbf{k}}\hat{r}_{\sigma}+\hat{p}_{\sigma}\hat{r}_{\mathbf{k}}\big)\,\overline{N}_{k}(\hat{a}_{\mathbf{k}\sigma}e^{i\mathbf{k}\cdot\mathbf{x}_{0}}-\hat{a}_{\mathbf{k}\sigma}^{\dagger}e^{-i\mathbf{k}\cdot\mathbf{x}_{0}}).
\end{align}

\section{Coupling spectral density in free space \label{sec:Coupling-spectral-density}}

Here we show the derivation of the coupling spectral density $J_{12}^{\mu\nu}(\omega)$
in free space, which is define by 
\begin{equation}
J_{12}^{\mu\nu}(\omega)=2\pi\sum_{\mathbf{k}\sigma}(g_{1,\mathbf{k}\sigma}^{\mu})^{*}g_{2,\mathbf{k}\sigma}^{\nu}\delta(\omega-\omega_{\mathbf{k}}),\qquad g_{i,\mathbf{k}\sigma}^{\mu}=-i\vec{m}_{i}^{\mu}\cdot(\hat{\mathrm{e}}_{\mathbf{k}}\times\hat{\mathrm{e}}_{\mathbf{k}\sigma})e^{i\mathbf{k}\cdot\mathbf{x}_{i}}\sqrt{\frac{\mu_{0}\hbar\omega_{k}}{2V}}.
\end{equation}
 Since the index $\mu$ only appears on $\vec{m}_{i}^{\mu}$ in $J_{12}^{\mu\nu}(\omega)$
to represent the transition/permanent dipole, hereafter we omit it
for simplicity. 

The summation over $\mathbf{k},\sigma$ is changed into integral by
\begin{equation}
\sum_{\mathbf{k},\sigma}[...]\quad\longrightarrow\quad\sum_{\sigma}\frac{V}{(2\pi)^{3}}\int d^{3}\mathbf{k}\,[...]=\sum_{\sigma}\frac{V}{(2\pi c)^{3}}\int\omega^{2}d\omega\int d\Omega\,[...]
\end{equation}
Thus the coupling spectral density $J_{12}(\omega)$ is given by (denoting
$\mathbf{r}:=\mathbf{x}_{2}-\mathbf{x}_{1}$)
\begin{align}
J_{12}(\omega) & =\frac{2\pi}{(2\pi c)^{3}}\,\frac{\mu_{0}\hbar\omega^{3}}{2}\int_{0}^{2\pi}d\varphi\int_{0}^{\pi}\sin\theta d\theta\,e^{i\mathbf{k}\cdot(\mathbf{x}_{2}-\mathbf{x}_{1})}\sum_{\sigma}[\vec{m}_{1}\cdot(\hat{\mathrm{e}}_{\mathbf{k}}\times\hat{\mathrm{e}}_{\mathbf{k}\sigma})][\vec{m}_{2}\cdot(\hat{\mathrm{e}}_{\mathbf{k}}\times\hat{\mathrm{e}}_{\mathbf{k}\sigma})]\nonumber \\
 & =\frac{1}{(2\pi)^{2}c^{3}}\,\frac{\mu_{0}\hbar\omega^{3}}{2}\int_{0}^{2\pi}d\varphi\int_{0}^{\pi}\sin\theta d\theta\,e^{i\mathbf{k}\cdot\mathbf{r}}\Big[\vec{m}_{1}\cdot\vec{m}_{2}-(\vec{m}_{1}\cdot\hat{\mathrm{e}}_{\mathbf{k}})(\vec{m}_{2}\cdot\hat{\mathrm{e}}_{\mathbf{k}})\Big].
\end{align}

To calculate the above integral, we use the vector $\mathbf{r}$ and
$\vec{m}_{1}$ to span a proper coordinate. We set $\hat{\mathrm{e}}_{z}:=\mathbf{r}/r:=\hat{\mathrm{e}}_{\mathbf{r}}$,
and $\hat{\mathrm{e}}_{y}:=\lambda^{\text{-}1}\mathbf{r}\times\vec{m}_{1}$,
where $\lambda=\sqrt{r^{2}\vec{m}_{1}^{2}+(\mathbf{r}\cdot\vec{m}_{1})^{2}}$
is a normalization factor, and $r:=|\mathbf{r}|$; then we have $\hat{\mathrm{e}}_{x}=\hat{\mathrm{e}}_{y}\times\hat{\mathrm{e}}_{z}=\lambda^{\text{-}1}[r\vec{m}_{1}-(\mathbf{r}\cdot\vec{m}_{1})\mathbf{r}/r]$.

With this basis $\hat{\mathrm{e}}_{x,y,z}$, the vectors in the above
integral can be written as 
\begin{align}
\hat{\mathrm{e}}_{\mathbf{k}} & =\sin\theta\cos\varphi\,\hat{\mathrm{e}}_{x}+\sin\theta\sin\varphi\,\hat{\mathrm{e}}_{y}+\cos\theta\,\hat{\mathrm{e}}_{z},\nonumber \\
\vec{m}_{1} & =\frac{\lambda}{r}\hat{\mathrm{e}}_{x}+(\vec{m}_{1}\cdot\hat{\mathrm{e}}_{\mathbf{r}})\hat{\mathrm{e}}_{z},\\
\vec{m}_{2} & =\sum_{i}(\vec{m}_{2}\cdot\hat{\mathrm{e}}_{i})\hat{\mathrm{e}}_{i}=\frac{r}{\lambda}[\vec{m}_{2}\cdot\vec{m}_{1}-(\vec{m}_{1}\cdot\hat{\mathrm{e}}_{\mathbf{r}})(\vec{m}_{2}\cdot\hat{\mathrm{e}}_{\mathbf{r}})]\hat{\mathrm{e}}_{x}+\frac{r}{\lambda}(\hat{\mathrm{e}}_{\mathbf{r}}\cdot\vec{m}_{1}\times\vec{m}_{2})\hat{\mathrm{e}}_{y}+(\vec{m}_{2}\cdot\hat{\mathrm{e}}_{\mathbf{r}})\hat{\mathrm{e}}_{z}.\nonumber 
\end{align}
Thus the products in the integral give
\begin{align}
\vec{m}_{1}\cdot\hat{\mathrm{e}}_{\mathbf{k}}= & \frac{\lambda}{r}\sin\theta\cos\varphi+(\vec{m}_{1}\cdot\hat{\mathrm{e}}_{\mathbf{r}})\cos\theta,\nonumber \\
\vec{m}_{2}\cdot\hat{\mathrm{e}}_{\mathbf{k}}= & \frac{r}{\lambda}(\vec{m}_{1}\times\hat{\mathrm{e}}_{\mathbf{r}})\cdot(\vec{m}_{2}\times\hat{\mathrm{e}}_{\mathbf{r}})\sin\theta\cos\varphi+\frac{r}{\lambda}(\hat{\mathrm{e}}_{\mathbf{r}}\cdot\vec{m}_{1}\times\vec{m}_{2})\sin\theta\sin\varphi+(\vec{m}_{2}\cdot\hat{\mathrm{e}}_{\mathbf{r}})\cos\theta.
\end{align}
In the product $(\vec{m}_{1}\cdot\hat{\mathrm{e}}_{\mathbf{k}})(\vec{m}_{2}\cdot\hat{\mathrm{e}}_{\mathbf{k}})$,
if we first integrate over $\varphi\in[0,2\pi]$, it turns out that
most terms vanish directly, and the remaining terms are
\begin{align}
\mathbf{I}_{0} & =\int_{0}^{2\pi}d\varphi\int_{0}^{\pi}\sin\theta d\theta\,e^{ikr\cos\theta}\vec{m}_{2}\cdot\vec{m}_{1}=4\pi\vec{m}_{2}\cdot\vec{m}_{1}\,\frac{\sin kr}{kr},\nonumber \\
\mathbf{I}_{1} & =\int_{0}^{2\pi}d\varphi\int_{0}^{\pi}\sin\theta d\theta\,e^{ikr\cos\theta}(\vec{m}_{1}\times\hat{\mathrm{e}}_{\mathbf{r}})\cdot(\vec{m}_{2}\times\hat{\mathrm{e}}_{\mathbf{r}})\sin^{2}\theta\cos^{2}\varphi=4\pi(\vec{m}_{1}\times\hat{\mathrm{e}}_{\mathbf{r}})\cdot(\vec{m}_{2}\times\hat{\mathrm{e}}_{\mathbf{r}})\big[\frac{\sin kr}{(kr)^{3}}-\frac{\cos kr}{(kr)^{2}}\big],\nonumber \\
\mathbf{I}_{2} & =\int_{0}^{2\pi}d\varphi\int_{0}^{\pi}\sin\theta d\theta\,e^{ikr\cos\theta}(\hat{m}_{1}\cdot\hat{\mathrm{e}}_{\mathbf{r}})(\hat{m}_{2}\cdot\hat{\mathrm{e}}_{\mathbf{r}})\cos^{2}\theta=4\pi(\vec{m}_{1}\cdot\hat{\mathrm{e}}_{\mathbf{r}})(\vec{m}_{2}\cdot\hat{\mathrm{e}}_{\mathbf{r}})\big[\frac{\sin kr}{kr}+\frac{2\cos kr}{(kr)^{2}}-\frac{2\sin kr}{(kr)^{3}}\big].
\end{align}
Therefore, we have
\begin{align}
J_{12}(\omega=ck)= & \frac{\mu_{0}\hbar k^{3}}{2\pi}\Big\{\vec{m}_{1}\cdot\vec{m}_{2}\frac{\sin kr}{kr}-(\vec{m}_{1}\times\hat{\mathrm{e}}_{\mathbf{r}})\cdot(\vec{m}_{2}\times\hat{\mathrm{e}}_{\mathbf{r}})\big[\frac{\sin kr}{(kr)^{3}}-\frac{\cos kr}{(kr)^{2}}\big]\nonumber \\
 & -(\vec{m}_{1}\cdot\hat{\mathrm{e}}_{\mathbf{r}})(\vec{m}_{2}\cdot\hat{\mathrm{e}}_{\mathbf{r}})\,\big[\frac{\sin kr}{kr}+\frac{2\cos kr}{(kr)^{2}}-\frac{2\sin kr}{(kr)^{3}}\big]\Big\}.
\end{align}
which is an odd function $J_{12}(-\omega)=-J_{12}(\omega)$, and we
also have $J_{12}(\omega)=J_{21}(\omega)$.

Notice that, using this coupling spectral density (proper indices
for $\text{\textsc{t}},\mathsf{e},\mathsf{g}$ should be added to
$\vec{m}_{i}$), the interaction strength of the permanent dipoles
is 
\begin{align}
\xi^{\text{\textsc{p}}}=-\int_{-\infty}^{\infty}\frac{d\omega}{2\pi\hbar}\,\frac{J_{12}(\omega)}{\omega} & =\frac{\mu_{0}}{r^{3}}\int_{-\infty}^{\infty}\frac{d(kr)}{2\pi}\,\frac{1}{kr}\cdot\frac{(kr)^{3}}{2\pi}\Big\{(\vec{m}_{1}\times\hat{\mathrm{e}}_{\mathbf{r}})\cdot(\vec{m}_{2}\times\hat{\mathrm{e}}_{\mathbf{r}})-2(\vec{m}_{1}\cdot\hat{\mathrm{e}}_{\mathbf{r}})(\vec{m}_{2}\cdot\hat{\mathrm{e}}_{\mathbf{r}})\Big\}\frac{\sin kr}{(kr)^{3}}\nonumber \\
 & =\frac{\mu_{0}}{4\pi r^{3}}\big[\vec{m}_{1}\cdot\vec{m}_{2}-3(\vec{m}_{1}\cdot\hat{\mathrm{e}}_{\mathbf{r}})(\vec{m}_{2}\cdot\hat{\mathrm{e}}_{\mathbf{r}})\big],\label{eq:apx-ksi-P}
\end{align}
which has exactly the same form with the classical dipole-dipole interaction.

On the other hand, the interaction strength between the transition
dipoles is given by 
\begin{align}
\xi^{\text{\textsc{t}}}=-\int_{-\infty}^{\infty}\frac{d\omega}{2\pi\hbar}\,\frac{J_{12}(\omega)}{\omega-\Omega}= & -\frac{\mu_{0}}{4\pi r^{3}}\int_{-\infty}^{\infty}\frac{dx}{\pi}\,\frac{x^{3}}{x-x_{\Omega}}\Big\{\vec{m}_{1}\cdot\vec{m}_{2}\frac{\sin x}{x}-(\vec{m}_{1}\times\hat{\mathrm{e}}_{\mathbf{r}})\cdot(\vec{m}_{2}\times\hat{\mathrm{e}}_{\mathbf{r}})\big[\frac{\sin x}{x^{3}}-\frac{\cos x}{x^{2}}\big]\nonumber \\
 & \qquad-(\vec{m}_{1}\cdot\hat{\mathrm{e}}_{\mathbf{r}})(\vec{m}_{2}\cdot\hat{\mathrm{e}}_{\mathbf{r}})\,\big[\frac{\sin x}{x}+\frac{2\cos x}{x^{2}}-\frac{2\sin x}{x^{3}}\big]\Big\}\nonumber \\
= & -\frac{\mu_{0}}{4\pi r^{3}}\Big\{\vec{m}_{1}\cdot\vec{m}_{2}x_{\Omega}^{2}\cos x_{\Omega}-(\vec{m}_{1}\times\hat{\mathrm{e}}_{\mathbf{r}})\cdot(\vec{m}_{2}\times\hat{\mathrm{e}}_{\mathbf{r}})\big[\cos x_{\Omega}-x_{\Omega}\sin x_{\Omega}\big]\nonumber \\
 & \qquad-(\vec{m}_{1}\cdot\hat{\mathrm{e}}_{\mathbf{r}})(\vec{m}_{2}\cdot\hat{\mathrm{e}}_{\mathbf{r}})\,\big[2x_{\Omega}\sin x_{\Omega}+x_{\Omega}^{2}\cos x_{\Omega}-2\cos x_{\Omega}\big]\Big\},
\end{align}
 where $x_{\Omega}:=r\Omega/c=2\pi r/\lambda$, and $\lambda$ is
the wavelength corresponding to the transition frequency $\Omega$.
In the short-distance limit $x_{\Omega}\rightarrow0$, the above interaction
strength $\xi^{\text{\textsc{t}}}$ returns to the same form as $\xi^{\text{\textsc{p}}}$
{[}Eq.\,(\ref{eq:apx-ksi-P}){]}, which is just the classical result. 

\end{widetext}

\bibliographystyle{apsrev4-1}
\bibliography{Refs}

\begin{thebibliography}{37}%
\makeatletter
\providecommand \@ifxundefined [1]{%
 \@ifx{#1\undefined}
}%
\providecommand \@ifnum [1]{%
 \ifnum #1\expandafter \@firstoftwo
 \else \expandafter \@secondoftwo
 \fi
}%
\providecommand \@ifx [1]{%
 \ifx #1\expandafter \@firstoftwo
 \else \expandafter \@secondoftwo
 \fi
}%
\providecommand \natexlab [1]{#1}%
\providecommand \enquote  [1]{``#1''}%
\providecommand \bibnamefont  [1]{#1}%
\providecommand \bibfnamefont [1]{#1}%
\providecommand \citenamefont [1]{#1}%
\providecommand \href@noop [0]{\@secondoftwo}%
\providecommand \href [0]{\begingroup \@sanitize@url \@href}%
\providecommand \@href[1]{\@@startlink{#1}\@@href}%
\providecommand \@@href[1]{\endgroup#1\@@endlink}%
\providecommand \@sanitize@url [0]{\catcode `\\12\catcode `\$12\catcode
  `\&12\catcode `\#12\catcode `\^12\catcode `\_12\catcode `\%12\relax}%
\providecommand \@@startlink[1]{}%
\providecommand \@@endlink[0]{}%
\providecommand \url  [0]{\begingroup\@sanitize@url \@url }%
\providecommand \@url [1]{\endgroup\@href {#1}{\urlprefix }}%
\providecommand \urlprefix  [0]{URL }%
\providecommand \Eprint [0]{\href }%
\providecommand \doibase [0]{http://dx.doi.org/}%
\providecommand \selectlanguage [0]{\@gobble}%
\providecommand \bibinfo  [0]{\@secondoftwo}%
\providecommand \bibfield  [0]{\@secondoftwo}%
\providecommand \translation [1]{[#1]}%
\providecommand \BibitemOpen [0]{}%
\providecommand \bibitemStop [0]{}%
\providecommand \bibitemNoStop [0]{.\EOS\space}%
\providecommand \EOS [0]{\spacefactor3000\relax}%
\providecommand \BibitemShut  [1]{\csname bibitem#1\endcsname}%
\let\auto@bib@innerbib\@empty
\bibitem [{\citenamefont {Liao}\ and\ \citenamefont
  {Zubairy}(2014)}]{liao_single-photon_2014}%
  \BibitemOpen
  \bibfield  {author} {\bibinfo {author} {\bibfnamefont {Z.}~\bibnamefont
  {Liao}}\ and\ \bibinfo {author} {\bibfnamefont {M.~S.}\ \bibnamefont
  {Zubairy}},\ }\href {\doibase 10.1103/PhysRevA.90.053805} {\bibfield
  {journal} {\bibinfo  {journal} {Phys. Rev. A}\ }\textbf {\bibinfo {volume}
  {90}},\ \bibinfo {pages} {053805} (\bibinfo {year} {2014})}\BibitemShut
  {NoStop}%
\bibitem [{\citenamefont {Liao}\ \emph {et~al.}(2015)\citenamefont {Liao},
  \citenamefont {Zeng}, \citenamefont {Zhu},\ and\ \citenamefont
  {Zubairy}}]{liao_single-photon_2015}%
  \BibitemOpen
  \bibfield  {author} {\bibinfo {author} {\bibfnamefont {Z.}~\bibnamefont
  {Liao}}, \bibinfo {author} {\bibfnamefont {X.}~\bibnamefont {Zeng}}, \bibinfo
  {author} {\bibfnamefont {S.-Y.}\ \bibnamefont {Zhu}}, \ and\ \bibinfo
  {author} {\bibfnamefont {M.~S.}\ \bibnamefont {Zubairy}},\ }\href {\doibase
  10.1103/PhysRevA.92.023806} {\bibfield  {journal} {\bibinfo  {journal} {Phys.
  Rev. A}\ }\textbf {\bibinfo {volume} {92}},\ \bibinfo {pages} {023806}
  (\bibinfo {year} {2015})}\BibitemShut {NoStop}%
\bibitem [{\citenamefont {Lambert}\ \emph {et~al.}(2016)\citenamefont
  {Lambert}, \citenamefont {Haigh}, \citenamefont {Langenfeld}, \citenamefont
  {Doherty},\ and\ \citenamefont {Ferguson}}]{lambert_cavity-mediated_2016}%
  \BibitemOpen
  \bibfield  {author} {\bibinfo {author} {\bibfnamefont {N.~J.}\ \bibnamefont
  {Lambert}}, \bibinfo {author} {\bibfnamefont {J.~A.}\ \bibnamefont {Haigh}},
  \bibinfo {author} {\bibfnamefont {S.}~\bibnamefont {Langenfeld}}, \bibinfo
  {author} {\bibfnamefont {A.~C.}\ \bibnamefont {Doherty}}, \ and\ \bibinfo
  {author} {\bibfnamefont {A.~J.}\ \bibnamefont {Ferguson}},\ }\href {\doibase
  10.1103/PhysRevA.93.021803} {\bibfield  {journal} {\bibinfo  {journal} {Phys.
  Rev. A}\ }\textbf {\bibinfo {volume} {93}},\ \bibinfo {pages} {021803}
  (\bibinfo {year} {2016})}\BibitemShut {NoStop}%
\bibitem [{\citenamefont {Baranov}\ \emph {et~al.}(2016)\citenamefont
  {Baranov}, \citenamefont {Savelev}, \citenamefont {Li}, \citenamefont
  {Krasnok},\ and\ \citenamefont {Al{\`u}}}]{baranov_magnetic_2016}%
  \BibitemOpen
  \bibfield  {author} {\bibinfo {author} {\bibfnamefont {D.~G.}\ \bibnamefont
  {Baranov}}, \bibinfo {author} {\bibfnamefont {R.~S.}\ \bibnamefont
  {Savelev}}, \bibinfo {author} {\bibfnamefont {S.~V.}\ \bibnamefont {Li}},
  \bibinfo {author} {\bibfnamefont {A.~E.}\ \bibnamefont {Krasnok}}, \ and\
  \bibinfo {author} {\bibfnamefont {A.}~\bibnamefont {Al{\`u}}},\ }\href
  {http://arxiv.org/abs/1610.02001} {\bibfield  {journal} {\bibinfo  {journal}
  {arXiv:1610.02001}\ } (\bibinfo {year} {2016})}\BibitemShut {NoStop}%
\bibitem [{\citenamefont {Liao}\ \emph {et~al.}(2016)\citenamefont {Liao},
  \citenamefont {Nha},\ and\ \citenamefont {Zubairy}}]{liao_dynamical_2016}%
  \BibitemOpen
  \bibfield  {author} {\bibinfo {author} {\bibfnamefont {Z.}~\bibnamefont
  {Liao}}, \bibinfo {author} {\bibfnamefont {H.}~\bibnamefont {Nha}}, \ and\
  \bibinfo {author} {\bibfnamefont {M.~S.}\ \bibnamefont {Zubairy}},\ }\href
  {\doibase 10.1103/PhysRevA.94.053842} {\bibfield  {journal} {\bibinfo
  {journal} {Phys. Rev. A}\ }\textbf {\bibinfo {volume} {94}},\ \bibinfo
  {pages} {053842} (\bibinfo {year} {2016})}\BibitemShut {NoStop}%
\bibitem [{\citenamefont {Shahmoon}\ and\ \citenamefont
  {Kurizki}(2013{\natexlab{a}})}]{shahmoon_nonradiative_2013}%
  \BibitemOpen
  \bibfield  {author} {\bibinfo {author} {\bibfnamefont {E.}~\bibnamefont
  {Shahmoon}}\ and\ \bibinfo {author} {\bibfnamefont {G.}~\bibnamefont
  {Kurizki}},\ }\href {\doibase 10.1103/PhysRevA.87.033831} {\bibfield
  {journal} {\bibinfo  {journal} {Phys. Rev. A}\ }\textbf {\bibinfo {volume}
  {87}},\ \bibinfo {pages} {033831} (\bibinfo {year}
  {2013}{\natexlab{a}})}\BibitemShut {NoStop}%
\bibitem [{\citenamefont {Shahmoon}\ and\ \citenamefont
  {Kurizki}(2013{\natexlab{b}})}]{shahmoon_dispersion_2013}%
  \BibitemOpen
  \bibfield  {author} {\bibinfo {author} {\bibfnamefont {E.}~\bibnamefont
  {Shahmoon}}\ and\ \bibinfo {author} {\bibfnamefont {G.}~\bibnamefont
  {Kurizki}},\ }\href {\doibase 10.1103/PhysRevA.87.062105} {\bibfield
  {journal} {\bibinfo  {journal} {Phys. Rev. A}\ }\textbf {\bibinfo {volume}
  {87}},\ \bibinfo {pages} {062105} (\bibinfo {year}
  {2013}{\natexlab{b}})}\BibitemShut {NoStop}%
\bibitem [{\citenamefont {Cai}\ \emph {et~al.}(2016)\citenamefont {Cai},
  \citenamefont {Wang}, \citenamefont {Svidzinsky}, \citenamefont {Zhu},\ and\
  \citenamefont {Scully}}]{cai_symmetry-protected_2016}%
  \BibitemOpen
  \bibfield  {author} {\bibinfo {author} {\bibfnamefont {H.}~\bibnamefont
  {Cai}}, \bibinfo {author} {\bibfnamefont {D.-W.}\ \bibnamefont {Wang}},
  \bibinfo {author} {\bibfnamefont {A.~A.}\ \bibnamefont {Svidzinsky}},
  \bibinfo {author} {\bibfnamefont {S.-Y.}\ \bibnamefont {Zhu}}, \ and\
  \bibinfo {author} {\bibfnamefont {M.~O.}\ \bibnamefont {Scully}},\ }\href
  {\doibase 10.1103/PhysRevA.93.053804} {\bibfield  {journal} {\bibinfo
  {journal} {Phys. Rev. A}\ }\textbf {\bibinfo {volume} {93}},\ \bibinfo
  {pages} {053804} (\bibinfo {year} {2016})}\BibitemShut {NoStop}%
\bibitem [{\citenamefont {Shahmoon}(2017)}]{shahmoon_casimir_2017}%
  \BibitemOpen
  \bibfield  {author} {\bibinfo {author} {\bibfnamefont {E.}~\bibnamefont
  {Shahmoon}},\ }\href {\doibase 10.1103/PhysRevA.95.062504} {\bibfield
  {journal} {\bibinfo  {journal} {Phys. Rev. A}\ }\textbf {\bibinfo {volume}
  {95}},\ \bibinfo {pages} {062504} (\bibinfo {year} {2017})}\BibitemShut
  {NoStop}%
\bibitem [{\citenamefont {Donaire}\ \emph {et~al.}(2017)\citenamefont
  {Donaire}, \citenamefont {Mu{\~n}oz-Casta{\~n}eda},\ and\ \citenamefont
  {Nieto}}]{donaire_dipole-dipole_2017}%
  \BibitemOpen
  \bibfield  {author} {\bibinfo {author} {\bibfnamefont {M.}~\bibnamefont
  {Donaire}}, \bibinfo {author} {\bibfnamefont {J.~M.}\ \bibnamefont
  {Mu{\~n}oz-Casta{\~n}eda}}, \ and\ \bibinfo {author} {\bibfnamefont {L.~M.}\
  \bibnamefont {Nieto}},\ }\href {\doibase 10.1103/PhysRevA.96.042714}
  {\bibfield  {journal} {\bibinfo  {journal} {Phys. Rev. A}\ }\textbf {\bibinfo
  {volume} {96}},\ \bibinfo {pages} {042714} (\bibinfo {year}
  {2017})}\BibitemShut {NoStop}%
\bibitem [{\citenamefont {Cortes}\ and\ \citenamefont
  {Jacob}(2017)}]{cortes_super-coulombic_2017}%
  \BibitemOpen
  \bibfield  {author} {\bibinfo {author} {\bibfnamefont {C.~L.}\ \bibnamefont
  {Cortes}}\ and\ \bibinfo {author} {\bibfnamefont {Z.}~\bibnamefont {Jacob}},\
  }\href {http://dx.doi.org/10.1038/ncomms14144} {\bibfield  {journal}
  {\bibinfo  {journal} {Nature Comm.}\ }\textbf {\bibinfo {volume} {8}},\
  \bibinfo {pages} {14144} (\bibinfo {year} {2017})}\BibitemShut {NoStop}%
\bibitem [{\citenamefont {Martinis}\ \emph {et~al.}(2005)\citenamefont
  {Martinis}, \citenamefont {Cooper}, \citenamefont {McDermott}, \citenamefont
  {Steffen}, \citenamefont {Ansmann}, \citenamefont {Osborn}, \citenamefont
  {Cicak}, \citenamefont {Oh}, \citenamefont {Pappas}, \citenamefont
  {Simmonds},\ and\ \citenamefont {Yu}}]{martinis_decoherence_2005}%
  \BibitemOpen
  \bibfield  {author} {\bibinfo {author} {\bibfnamefont {J.~M.}\ \bibnamefont
  {Martinis}}, \bibinfo {author} {\bibfnamefont {K.~B.}\ \bibnamefont
  {Cooper}}, \bibinfo {author} {\bibfnamefont {R.}~\bibnamefont {McDermott}},
  \bibinfo {author} {\bibfnamefont {M.}~\bibnamefont {Steffen}}, \bibinfo
  {author} {\bibfnamefont {M.}~\bibnamefont {Ansmann}}, \bibinfo {author}
  {\bibfnamefont {K.~D.}\ \bibnamefont {Osborn}}, \bibinfo {author}
  {\bibfnamefont {K.}~\bibnamefont {Cicak}}, \bibinfo {author} {\bibfnamefont
  {S.}~\bibnamefont {Oh}}, \bibinfo {author} {\bibfnamefont {D.~P.}\
  \bibnamefont {Pappas}}, \bibinfo {author} {\bibfnamefont {R.~W.}\
  \bibnamefont {Simmonds}}, \ and\ \bibinfo {author} {\bibfnamefont {C.~C.}\
  \bibnamefont {Yu}},\ }\href {\doibase 10.1103/PhysRevLett.95.210503}
  {\bibfield  {journal} {\bibinfo  {journal} {Phys. Rev. Lett.}\ }\textbf
  {\bibinfo {volume} {95}},\ \bibinfo {pages} {210503} (\bibinfo {year}
  {2005})}\BibitemShut {NoStop}%
\bibitem [{\citenamefont {Paik}\ \emph {et~al.}(2011)\citenamefont {Paik},
  \citenamefont {Schuster}, \citenamefont {Bishop}, \citenamefont {Kirchmair},
  \citenamefont {Catelani}, \citenamefont {Sears}, \citenamefont {Johnson},
  \citenamefont {Reagor}, \citenamefont {Frunzio}, \citenamefont {Glazman},
  \citenamefont {Girvin}, \citenamefont {Devoret},\ and\ \citenamefont
  {Schoelkopf}}]{paik_observation_2011}%
  \BibitemOpen
  \bibfield  {author} {\bibinfo {author} {\bibfnamefont {H.}~\bibnamefont
  {Paik}}, \bibinfo {author} {\bibfnamefont {D.~I.}\ \bibnamefont {Schuster}},
  \bibinfo {author} {\bibfnamefont {L.~S.}\ \bibnamefont {Bishop}}, \bibinfo
  {author} {\bibfnamefont {G.}~\bibnamefont {Kirchmair}}, \bibinfo {author}
  {\bibfnamefont {G.}~\bibnamefont {Catelani}}, \bibinfo {author}
  {\bibfnamefont {A.~P.}\ \bibnamefont {Sears}}, \bibinfo {author}
  {\bibfnamefont {B.~R.}\ \bibnamefont {Johnson}}, \bibinfo {author}
  {\bibfnamefont {M.~J.}\ \bibnamefont {Reagor}}, \bibinfo {author}
  {\bibfnamefont {L.}~\bibnamefont {Frunzio}}, \bibinfo {author} {\bibfnamefont
  {L.~I.}\ \bibnamefont {Glazman}}, \bibinfo {author} {\bibfnamefont {S.~M.}\
  \bibnamefont {Girvin}}, \bibinfo {author} {\bibfnamefont {M.~H.}\
  \bibnamefont {Devoret}}, \ and\ \bibinfo {author} {\bibfnamefont {R.~J.}\
  \bibnamefont {Schoelkopf}},\ }\href {\doibase 10.1103/PhysRevLett.107.240501}
  {\bibfield  {journal} {\bibinfo  {journal} {Phys. Rev. Lett.}\ }\textbf
  {\bibinfo {volume} {107}},\ \bibinfo {pages} {240501} (\bibinfo {year}
  {2011})}\BibitemShut {NoStop}%
\bibitem [{\citenamefont {Rigetti}\ \emph {et~al.}(2012)\citenamefont
  {Rigetti}, \citenamefont {Gambetta}, \citenamefont {Poletto}, \citenamefont
  {Plourde}, \citenamefont {Chow}, \citenamefont {C{\'o}rcoles}, \citenamefont
  {Smolin}, \citenamefont {Merkel}, \citenamefont {Rozen}, \citenamefont
  {Keefe}, \citenamefont {Rothwell}, \citenamefont {Ketchen},\ and\
  \citenamefont {Steffen}}]{rigetti_superconducting_2012}%
  \BibitemOpen
  \bibfield  {author} {\bibinfo {author} {\bibfnamefont {C.}~\bibnamefont
  {Rigetti}}, \bibinfo {author} {\bibfnamefont {J.~M.}\ \bibnamefont
  {Gambetta}}, \bibinfo {author} {\bibfnamefont {S.}~\bibnamefont {Poletto}},
  \bibinfo {author} {\bibfnamefont {B.~L.~T.}\ \bibnamefont {Plourde}},
  \bibinfo {author} {\bibfnamefont {J.~M.}\ \bibnamefont {Chow}}, \bibinfo
  {author} {\bibfnamefont {A.~D.}\ \bibnamefont {C{\'o}rcoles}}, \bibinfo
  {author} {\bibfnamefont {J.~A.}\ \bibnamefont {Smolin}}, \bibinfo {author}
  {\bibfnamefont {S.~T.}\ \bibnamefont {Merkel}}, \bibinfo {author}
  {\bibfnamefont {J.~R.}\ \bibnamefont {Rozen}}, \bibinfo {author}
  {\bibfnamefont {G.~A.}\ \bibnamefont {Keefe}}, \bibinfo {author}
  {\bibfnamefont {M.~B.}\ \bibnamefont {Rothwell}}, \bibinfo {author}
  {\bibfnamefont {M.~B.}\ \bibnamefont {Ketchen}}, \ and\ \bibinfo {author}
  {\bibfnamefont {M.}~\bibnamefont {Steffen}},\ }\href {\doibase
  10.1103/PhysRevB.86.100506} {\bibfield  {journal} {\bibinfo  {journal}
  {Physical Review B}\ }\textbf {\bibinfo {volume} {86}},\ \bibinfo {pages}
  {100506} (\bibinfo {year} {2012})}\BibitemShut {NoStop}%
\bibitem [{\citenamefont {Lisenfeld}\ \emph {et~al.}(2016)\citenamefont
  {Lisenfeld}, \citenamefont {Bilmes}, \citenamefont {Matityahu}, \citenamefont
  {Zanker}, \citenamefont {Marthaler}, \citenamefont {Schechter}, \citenamefont
  {Sch{\"o}n}, \citenamefont {Shnirman}, \citenamefont {Weiss},\ and\
  \citenamefont {Ustinov}}]{lisenfeld_decoherence_2016}%
  \BibitemOpen
  \bibfield  {author} {\bibinfo {author} {\bibfnamefont {J.}~\bibnamefont
  {Lisenfeld}}, \bibinfo {author} {\bibfnamefont {A.}~\bibnamefont {Bilmes}},
  \bibinfo {author} {\bibfnamefont {S.}~\bibnamefont {Matityahu}}, \bibinfo
  {author} {\bibfnamefont {S.}~\bibnamefont {Zanker}}, \bibinfo {author}
  {\bibfnamefont {M.}~\bibnamefont {Marthaler}}, \bibinfo {author}
  {\bibfnamefont {M.}~\bibnamefont {Schechter}}, \bibinfo {author}
  {\bibfnamefont {G.}~\bibnamefont {Sch{\"o}n}}, \bibinfo {author}
  {\bibfnamefont {A.}~\bibnamefont {Shnirman}}, \bibinfo {author}
  {\bibfnamefont {G.}~\bibnamefont {Weiss}}, \ and\ \bibinfo {author}
  {\bibfnamefont {A.~V.}\ \bibnamefont {Ustinov}},\ }\href {\doibase
  10.1038/srep23786} {\bibfield  {journal} {\bibinfo  {journal} {Sci. Rep.}\
  }\textbf {\bibinfo {volume} {6}},\ \bibinfo {pages} {srep23786} (\bibinfo
  {year} {2016})}\BibitemShut {NoStop}%
\bibitem [{\citenamefont {Doherty}\ \emph {et~al.}(2013)\citenamefont
  {Doherty}, \citenamefont {Manson}, \citenamefont {Delaney}, \citenamefont
  {Jelezko}, \citenamefont {Wrachtrup},\ and\ \citenamefont
  {Hollenberg}}]{doherty_nitrogen-vacancy_2013}%
  \BibitemOpen
  \bibfield  {author} {\bibinfo {author} {\bibfnamefont {M.~W.}\ \bibnamefont
  {Doherty}}, \bibinfo {author} {\bibfnamefont {N.~B.}\ \bibnamefont {Manson}},
  \bibinfo {author} {\bibfnamefont {P.}~\bibnamefont {Delaney}}, \bibinfo
  {author} {\bibfnamefont {F.}~\bibnamefont {Jelezko}}, \bibinfo {author}
  {\bibfnamefont {J.}~\bibnamefont {Wrachtrup}}, \ and\ \bibinfo {author}
  {\bibfnamefont {L.~C.~L.}\ \bibnamefont {Hollenberg}},\ }\href {\doibase
  10.1016/j.physrep.2013.02.001} {\bibfield  {journal} {\bibinfo  {journal}
  {Phys. Rep.}\ }\textbf {\bibinfo {volume} {528}},\ \bibinfo {pages} {1}
  (\bibinfo {year} {2013})}\BibitemShut {NoStop}%
\bibitem [{\citenamefont {Zhao}\ \emph {et~al.}(2012)\citenamefont {Zhao},
  \citenamefont {Ho},\ and\ \citenamefont {Liu}}]{zhao_decoherence_2012}%
  \BibitemOpen
  \bibfield  {author} {\bibinfo {author} {\bibfnamefont {N.}~\bibnamefont
  {Zhao}}, \bibinfo {author} {\bibfnamefont {S.-W.}\ \bibnamefont {Ho}}, \ and\
  \bibinfo {author} {\bibfnamefont {R.-B.}\ \bibnamefont {Liu}},\ }\href
  {\doibase 10.1103/PhysRevB.85.115303} {\bibfield  {journal} {\bibinfo
  {journal} {Phys. Rev. B}\ }\textbf {\bibinfo {volume} {85}},\ \bibinfo
  {pages} {115303} (\bibinfo {year} {2012})}\BibitemShut {NoStop}%
\bibitem [{\citenamefont {Yang}\ \emph {et~al.}(2010)\citenamefont {Yang},
  \citenamefont {Xu}, \citenamefont {Song},\ and\ \citenamefont
  {Sun}}]{yang_dimerization-assisted_2010}%
  \BibitemOpen
  \bibfield  {author} {\bibinfo {author} {\bibfnamefont {S.}~\bibnamefont
  {Yang}}, \bibinfo {author} {\bibfnamefont {D.~Z.}\ \bibnamefont {Xu}},
  \bibinfo {author} {\bibfnamefont {Z.}~\bibnamefont {Song}}, \ and\ \bibinfo
  {author} {\bibfnamefont {C.~P.}\ \bibnamefont {Sun}},\ }\href {\doibase
  doi:10.1063/1.3435213} {\bibfield  {journal} {\bibinfo  {journal} {J. Chem.
  Phys.}\ }\textbf {\bibinfo {volume} {132}},\ \bibinfo {pages} {234501}
  (\bibinfo {year} {2010})}\BibitemShut {NoStop}%
\bibitem [{\citenamefont {El-Ganainy}\ and\ \citenamefont
  {John}(2013)}]{el-ganainy_resonant_2013}%
  \BibitemOpen
  \bibfield  {author} {\bibinfo {author} {\bibfnamefont {R.}~\bibnamefont
  {El-Ganainy}}\ and\ \bibinfo {author} {\bibfnamefont {S.}~\bibnamefont
  {John}},\ }\href {\doibase 10.1088/1367-2630/15/8/083033} {\bibfield
  {journal} {\bibinfo  {journal} {New J. Phys.}\ }\textbf {\bibinfo {volume}
  {15}},\ \bibinfo {pages} {083033} (\bibinfo {year} {2013})}\BibitemShut
  {NoStop}%
\bibitem [{\citenamefont {Dong}\ \emph {et~al.}(2016)\citenamefont {Dong},
  \citenamefont {Li}, \citenamefont {Yi}, \citenamefont {Agarwal},\ and\
  \citenamefont {Scully}}]{dong_photon-blockade_2016}%
  \BibitemOpen
  \bibfield  {author} {\bibinfo {author} {\bibfnamefont {H.}~\bibnamefont
  {Dong}}, \bibinfo {author} {\bibfnamefont {S.-W.}\ \bibnamefont {Li}},
  \bibinfo {author} {\bibfnamefont {Z.}~\bibnamefont {Yi}}, \bibinfo {author}
  {\bibfnamefont {G.~S.}\ \bibnamefont {Agarwal}}, \ and\ \bibinfo {author}
  {\bibfnamefont {M.~O.}\ \bibnamefont {Scully}},\ }\href
  {http://arxiv.org/abs/1608.04364} {\bibfield  {journal} {\bibinfo  {journal}
  {arXiv:1608.04364}\ } (\bibinfo {year} {2016})}\BibitemShut {NoStop}%
\bibitem [{\citenamefont {Jackson}(1998)}]{jackson_classical_1998}%
  \BibitemOpen
  \bibfield  {author} {\bibinfo {author} {\bibfnamefont {J.~D.}\ \bibnamefont
  {Jackson}},\ }\href@noop {} {\emph {\bibinfo {title} {Classical
  {Electrodynamics} {Third} {Edition}}}},\ \bibinfo {edition} {3rd}\ ed.\
  (\bibinfo  {publisher} {Wiley},\ \bibinfo {address} {New York},\ \bibinfo
  {year} {1998})\BibitemShut {NoStop}%
\bibitem [{\citenamefont {Lyuboshitz}(1968)}]{lyuboshitz_resonance_1968}%
  \BibitemOpen
  \bibfield  {author} {\bibinfo {author} {\bibfnamefont {V.~L.}\ \bibnamefont
  {Lyuboshitz}},\ }\href@noop {} {\bibfield  {journal} {\bibinfo  {journal}
  {Sov. Phys. JETP}\ }\textbf {\bibinfo {volume} {26}},\ \bibinfo {pages} {937}
  (\bibinfo {year} {1968})}\BibitemShut {NoStop}%
\bibitem [{\citenamefont {Lehmberg}(1970)}]{lehmberg_radiation_1970}%
  \BibitemOpen
  \bibfield  {author} {\bibinfo {author} {\bibfnamefont {R.~H.}\ \bibnamefont
  {Lehmberg}},\ }\href {\doibase 10.1103/PhysRevA.2.883} {\bibfield  {journal}
  {\bibinfo  {journal} {Phys. Rev. A}\ }\textbf {\bibinfo {volume} {2}},\
  \bibinfo {pages} {883} (\bibinfo {year} {1970})}\BibitemShut {NoStop}%
\bibitem [{\citenamefont {Ficek}\ \emph {et~al.}(1987)\citenamefont {Ficek},
  \citenamefont {Tana{\'s}},\ and\ \citenamefont
  {Kielich}}]{ficek_quantum_1987}%
  \BibitemOpen
  \bibfield  {author} {\bibinfo {author} {\bibfnamefont {Z.}~\bibnamefont
  {Ficek}}, \bibinfo {author} {\bibfnamefont {R.}~\bibnamefont {Tana{\'s}}}, \
  and\ \bibinfo {author} {\bibfnamefont {S.}~\bibnamefont {Kielich}},\ }\href
  {\doibase 10.1016/0378-4371(87)90280-9} {\bibfield  {journal} {\bibinfo
  {journal} {Physica A}\ }\textbf {\bibinfo {volume} {146}},\ \bibinfo {pages}
  {452} (\bibinfo {year} {1987})}\BibitemShut {NoStop}%
\bibitem [{\citenamefont {Agarwal}(1974)}]{agarwal_quantum_1974}%
  \BibitemOpen
  \bibfield  {author} {\bibinfo {author} {\bibfnamefont {G.~S.}\ \bibnamefont
  {Agarwal}},\ }\href@noop {} {\emph {\bibinfo {title} {Quantum statistical
  theories of spontaneous emission and their relation to other approaches}}}\
  (\bibinfo  {publisher} {Springer},\ \bibinfo {year} {1974})\BibitemShut
  {NoStop}%
\bibitem [{\citenamefont {Ficek}\ and\ \citenamefont
  {Swain}(2005)}]{ficek_quantum_2005}%
  \BibitemOpen
  \bibfield  {author} {\bibinfo {author} {\bibfnamefont {Z.}~\bibnamefont
  {Ficek}}\ and\ \bibinfo {author} {\bibfnamefont {S.}~\bibnamefont {Swain}},\
  }\href@noop {} {\emph {\bibinfo {title} {Quantum {Interference} and
  {Coherence}: {Theory} and {Experiments}}}},\ \bibinfo {edition} {1st}\ ed.\
  (\bibinfo  {publisher} {Springer},\ \bibinfo {address} {Berlin},\ \bibinfo
  {year} {2005})\BibitemShut {NoStop}%
\bibitem [{\citenamefont {Dicke}(1954)}]{dicke_coherence_1954}%
  \BibitemOpen
  \bibfield  {author} {\bibinfo {author} {\bibfnamefont {R.~H.}\ \bibnamefont
  {Dicke}},\ }\href {\doibase 10.1103/PhysRev.93.99} {\bibfield  {journal}
  {\bibinfo  {journal} {Phys. Rev.}\ }\textbf {\bibinfo {volume} {93}},\
  \bibinfo {pages} {99} (\bibinfo {year} {1954})}\BibitemShut {NoStop}%
\bibitem [{\citenamefont {Cohen-Tannoudji}\ \emph {et~al.}(1989)\citenamefont
  {Cohen-Tannoudji}, \citenamefont {Dupont-Roc},\ and\ \citenamefont
  {Grynberg}}]{cohen-tannoudji_photons_1989}%
  \BibitemOpen
  \bibfield  {author} {\bibinfo {author} {\bibfnamefont {C.}~\bibnamefont
  {Cohen-Tannoudji}}, \bibinfo {author} {\bibfnamefont {J.}~\bibnamefont
  {Dupont-Roc}}, \ and\ \bibinfo {author} {\bibfnamefont {G.}~\bibnamefont
  {Grynberg}},\ }\href@noop {} {\emph {\bibinfo {title} {Photons and {Atoms}:
  {Introduction} to {Quantum} {Electrodynamics}}}},\ \bibinfo {edition} {1st}\
  ed.\ (\bibinfo  {publisher} {Wiley-VCH},\ \bibinfo {address} {New York},\
  \bibinfo {year} {1989})\BibitemShut {NoStop}%
\bibitem [{\citenamefont {Ryder}(1996)}]{ryder_quantum_1996}%
  \BibitemOpen
  \bibfield  {author} {\bibinfo {author} {\bibfnamefont {L.~H.}\ \bibnamefont
  {Ryder}},\ }\href@noop {} {\emph {\bibinfo {title} {Quantum {Field}
  {Theory}}}},\ \bibinfo {edition} {2nd}\ ed.\ (\bibinfo  {publisher}
  {Cambridge University Press},\ \bibinfo {address} {Cambridge ; New York},\
  \bibinfo {year} {1996})\BibitemShut {NoStop}%
\bibitem [{\citenamefont {Purcell}\ \emph {et~al.}(1946)\citenamefont
  {Purcell}, \citenamefont {Torrey},\ and\ \citenamefont
  {Pound}}]{purcell_resonance_1946}%
  \BibitemOpen
  \bibfield  {author} {\bibinfo {author} {\bibfnamefont {E.~M.}\ \bibnamefont
  {Purcell}}, \bibinfo {author} {\bibfnamefont {H.~C.}\ \bibnamefont {Torrey}},
  \ and\ \bibinfo {author} {\bibfnamefont {R.~V.}\ \bibnamefont {Pound}},\
  }\href {\doibase 10.1103/PhysRev.69.37} {\bibfield  {journal} {\bibinfo
  {journal} {Phys. Rev.}\ }\textbf {\bibinfo {volume} {69}},\ \bibinfo {pages}
  {37} (\bibinfo {year} {1946})}\BibitemShut {NoStop}%
\bibitem [{\citenamefont {Scully}\ and\ \citenamefont
  {Zubairy}(1997)}]{scully_quantum_1997}%
  \BibitemOpen
  \bibfield  {author} {\bibinfo {author} {\bibfnamefont {M.~O.}\ \bibnamefont
  {Scully}}\ and\ \bibinfo {author} {\bibfnamefont {M.~S.}\ \bibnamefont
  {Zubairy}},\ }\href@noop {} {\emph {\bibinfo {title} {Quantum optics}}}\
  (\bibinfo  {publisher} {Cambridge university press},\ \bibinfo {year}
  {1997})\BibitemShut {NoStop}%
\bibitem [{\citenamefont {Weinberg}(2012)}]{weinberg_lectures_2012}%
  \BibitemOpen
  \bibfield  {author} {\bibinfo {author} {\bibfnamefont {S.}~\bibnamefont
  {Weinberg}},\ }\href@noop {} {\emph {\bibinfo {title} {Lectures on quantum
  mechanics}}}\ (\bibinfo  {publisher} {Cambridge University Press},\ \bibinfo
  {year} {2012})\BibitemShut {NoStop}%
\bibitem [{\citenamefont {Breuer}\ and\ \citenamefont
  {Petruccione}(2002)}]{breuer_theory_2002}%
  \BibitemOpen
  \bibfield  {author} {\bibinfo {author} {\bibfnamefont {H.}~\bibnamefont
  {Breuer}}\ and\ \bibinfo {author} {\bibfnamefont {F.}~\bibnamefont
  {Petruccione}},\ }\href@noop {} {\emph {\bibinfo {title} {The theory of open
  quantum systems}}}\ (\bibinfo  {publisher} {Oxford University Press},\
  \bibinfo {year} {2002})\BibitemShut {NoStop}%
\bibitem [{\citenamefont {Li}\ \emph {et~al.}(2015)\citenamefont {Li},
  \citenamefont {Cai},\ and\ \citenamefont {Sun}}]{li_steady_2015}%
  \BibitemOpen
  \bibfield  {author} {\bibinfo {author} {\bibfnamefont {S.-W.}\ \bibnamefont
  {Li}}, \bibinfo {author} {\bibfnamefont {C.~Y.}\ \bibnamefont {Cai}}, \ and\
  \bibinfo {author} {\bibfnamefont {C.~P.}\ \bibnamefont {Sun}},\ }\href
  {\doibase 10.1016/j.aop.2015.05.004} {\bibfield  {journal} {\bibinfo
  {journal} {Ann. Phys.}\ }\textbf {\bibinfo {volume} {360}},\ \bibinfo {pages}
  {19} (\bibinfo {year} {2015})},\ \bibinfo {note} {arXiv:
  1407.4290}\BibitemShut {NoStop}%
\bibitem [{\citenamefont {Li}\ \emph {et~al.}(2014)\citenamefont {Li},
  \citenamefont {Yang},\ and\ \citenamefont {Sun}}]{li_long-term_2014}%
  \BibitemOpen
  \bibfield  {author} {\bibinfo {author} {\bibfnamefont {S.-W.}\ \bibnamefont
  {Li}}, \bibinfo {author} {\bibfnamefont {L.-P.}\ \bibnamefont {Yang}}, \ and\
  \bibinfo {author} {\bibfnamefont {C.-P.}\ \bibnamefont {Sun}},\ }\href
  {\doibase 10.1140/epjd/e2014-40659-8} {\bibfield  {journal} {\bibinfo
  {journal} {Eur. Phys. J. D}\ }\textbf {\bibinfo {volume} {68}},\ \bibinfo
  {pages} {45} (\bibinfo {year} {2014})},\ \bibinfo {note}
  {arXiv:1303.1266}\BibitemShut {NoStop}%
\bibitem [{\citenamefont {Gardiner}\ and\ \citenamefont
  {Zoller}(2004)}]{gardiner_quantum_2004}%
  \BibitemOpen
  \bibfield  {author} {\bibinfo {author} {\bibfnamefont {C.}~\bibnamefont
  {Gardiner}}\ and\ \bibinfo {author} {\bibfnamefont {P.}~\bibnamefont
  {Zoller}},\ }\href@noop {} {\emph {\bibinfo {title} {Quantum noise}}},\
  Vol.~\bibinfo {volume} {56}\ (\bibinfo  {publisher} {Springer},\ \bibinfo
  {year} {2004})\BibitemShut {NoStop}%
\bibitem [{\citenamefont {Li}(2017)}]{li_production_2017}%
  \BibitemOpen
  \bibfield  {author} {\bibinfo {author} {\bibfnamefont {S.-W.}\ \bibnamefont
  {Li}},\ }\href {\doibase 10.1103/PhysRevE.96.012139} {\bibfield  {journal}
  {\bibinfo  {journal} {Phys. Rev. E}\ }\textbf {\bibinfo {volume} {96}},\
  \bibinfo {pages} {012139} (\bibinfo {year} {2017})},\ \bibinfo {note} {arXiv:
  1612.03884}\BibitemShut {NoStop}%
\end{thebibliography}%

\end{document}